\documentclass[twoside,twocolumn,9pt]{article}
\usepackage{extsizes}
\usepackage[super,sort&compress,comma]{natbib} 
\usepackage[version=3]{mhchem}
\usepackage[left=1.5cm, right=1.5cm, top=1.785cm, bottom=2.0cm]{geometry}
\usepackage{balance}
\usepackage{physics}

\usepackage{newtxtext}
\usepackage{newtxmath}

\usepackage{sectsty}
\usepackage{graphicx} 
\usepackage{lastpage}
\usepackage{amssymb}
\usepackage[format=plain,justification=justified,singlelinecheck=false,font={stretch=1.125,small,sf},labelfont=bf,labelsep=space]{caption}
\usepackage{float}
\usepackage{fancyhdr}
\usepackage{fnpos}
\usepackage[english]{babel}
\addto{\captionsenglish}{%
  
}
\usepackage{array}
\usepackage{droidsans}
\usepackage{charter}
\usepackage[T1]{fontenc}
\usepackage[usenames,dvipsnames]{xcolor}
\usepackage{setspace}
\usepackage[compact]{titlesec}
\usepackage{hyperref}

\usepackage{epstopdf}

\definecolor{cream}{RGB}{222,217,201}

\newcommand*{\rom}[1]{\expandafter\@slowromancap\romannumeral #1@}

\renewcommand{\uparrow}{\mathrm{f}}
\renewcommand{\downarrow}{\mathrm{e}}

\newcommand*{\citen}[1]{%
  \begingroup
    \romannumeral-`\x 
    \setcitestyle{numbers}%
    \cite{#1}%
  \endgroup   
}
\newcommand{\angstrom}{\mbox{\normalfont\AA}}

\begin{document}

\pagestyle{fancy}
\thispagestyle{plain}
\fancypagestyle{plain}{
\renewcommand{\headrulewidth}{0pt}
}

\makeFNbottom
\makeatletter
\renewcommand\LARGE{\@setfontsize\LARGE{15pt}{17}}
\renewcommand\Large{\@setfontsize\Large{12pt}{14}}
\renewcommand\large{\@setfontsize\large{10pt}{12}}
\renewcommand\footnotesize{\@setfontsize\footnotesize{7pt}{10}}
\makeatother

\renewcommand{\thefootnote}{\fnsymbol{footnote}}
\renewcommand\footnoterule{\vspace*{1pt}%
\color{cream}\hrule width 3.5in height 0.4pt \color{black}\vspace*{5pt}} 
\setcounter{secnumdepth}{5}

\makeatletter 
\renewcommand\@biblabel[1]{#1}            
\renewcommand\@makefntext[1]%
{\noindent\makebox[0pt][r]{\@thefnmark\,}#1}
\makeatother 
\renewcommand{\figurename}{\small{Fig.}~}
\sectionfont{\sffamily\Large}
\subsectionfont{\normalsize}
\subsubsectionfont{\bf}
\setstretch{1.125} 
\setlength{\skip\footins}{0.8cm}
\setlength{\footnotesep}{0.25cm}
\setlength{\jot}{10pt}
\titlespacing*{\section}{0pt}{4pt}{4pt}
\titlespacing*{\subsection}{0pt}{15pt}{1pt}

\fancyfoot{}
\fancyfoot[RO]{\footnotesize{\sffamily{1--\pageref{LastPage} ~\textbar  \hspace{2pt}\thepage}}}
\fancyfoot[LE]{\footnotesize{\sffamily{\thepage~\textbar\hspace{4.65cm} 1--\pageref{LastPage}}}}
\fancyhead{}
\renewcommand{\headrulewidth}{0pt} 
\renewcommand{\footrulewidth}{0pt}
\setlength{\arrayrulewidth}{1pt}
\setlength{\columnsep}{6.5mm}
\setlength\bibsep{1pt}

\makeatletter 
\newlength{\figrulesep} 
\setlength{\figrulesep}{0.5\textfloatsep} 

\newcommand{\topfigrule}{\vspace*{-1pt}%
\noindent{\color{cream}\rule[-\figrulesep]{\columnwidth}{1.5pt}} }

\newcommand{\botfigrule}{\vspace*{-2pt}%
\noindent{\color{cream}\rule[\figrulesep]{\columnwidth}{1.5pt}} }

\newcommand{\dblfigrule}{\vspace*{-1pt}%
\noindent{\color{cream}\rule[-\figrulesep]{\textwidth}{1.5pt}} }

\makeatother

\twocolumn[
  \begin{@twocolumnfalse}
  \noindent\hspace{0.075\linewidth}\begin{minipage}{0.85\textwidth}

\noindent\LARGE{\textbf{Dipole-phonon quantum logic with alkaline-earth monoxide and monosulfide cations}}
\vspace{0.3cm}

\noindent\large{Michael Mills,\textit{$^{a}$} Hao Wu,\textit{$^{a}$} Evan C. Reed,\textit{$^{b}$} Lu Qi,\textit{$^{b}$} Kenneth R. Brown,\textit{$^{b}$} Christian Schneider,\textit{$^{a}$} Michael C. Heaven,\textit{$^{c}$} Wesley C. Campbell,\textit{$^{ad}$} and Eric R. Hudson\textit{$^{ad}$}} \\

\textit{$^{a}$~Department of Physics and Astronomy, University of California, Los Angeles, California 90095, USA.}\\
\textit{$^{b}$~Departments of Electrical and Computer Engineering, Chemistry, and Physics, Duke University, Durham, North Carolina 27708, USA.}\\
\textit{$^{c}$~Department of Chemistry, Emory University, Atlanta, Georgia 30322, USA.}\\
\textit{$^{d}$~Center for Quantum Science and Engineering, University of California, Los Angeles, California 90095, USA.}

\begin{center}
(Dated: \today)
\end{center}

\vspace{0.3cm}

\normalsize{
\hspace{0.2cm} Dipole-phonon quantum logic (DPQL) leverages the interaction between polar molecular ions and the motional modes of a trapped-ion Coulomb crystal to provide a potentially scalable route to quantum information science.
Here, we study a class of candidate molecular ions for DPQL, the cationic alkaline-earth monoxides and monosulfides, which possess suitable structure for DPQL and can be produced in existing atomic ion experiments with little additional complexity.
We present calculations of DPQL operations for one of these molecules, CaO$^+$, and discuss progress towards experimental realization. 
We also further develop the theory of DPQL to include state preparation and measurement and entanglement of multiple molecular ions.
}

\end{minipage}
 \end{@twocolumnfalse} \vspace{0.6cm}

  ]

\renewcommand*\rmdefault{bch}\normalfont\upshape
\rmfamily
\section*{}
\vspace{-1cm}




\section{Introduction}
Trapped-ion qubits have demonstrated the highest-fidelity quantum operations of all quantum systems~\cite{Ballance16,Harty14,Gaebler16}.
As such, they are promising candidates for a scalable quantum information platform if the challenge of integrating and scaling the associated technology can be met. 
Chief among these challenges is this integration of the lasers required not only for cooling of the ions, but typically for manipulation of the qubits.
Currently, two main approaches to this problem are being pursued.
First, integrated photonics could provide a scalable means to deliver the requisite lasers~\cite{Bruzewicz2019,Karan2020} if the capabilities demonstrated in silicon photonics can be extended to materials compatible with the visible and ultraviolet wavelengths necessary for atomic ion qubits~\cite{Mehta16}. 
Second, several schemes for laser-less manipulation of atomic ion qubits are being explored, which involve microwave fields paired with either a strong static magnetic field gradient \cite{Mintert2001,Johanning2009,Khromova2012}, a microwave magnetic field gradient \cite{Ospelkaus2008,Ospelkaus2011,Brown2011}, microwave dressed states \cite{Timoney2011}, or a magnetic field gradient oscillating near a motional mode frequency \cite{Srinivas2019,Sutherland2019}.
Both integrated optics and microwave control require advances in ion trap fabrication to be truly scalable.

A recent proposal~\cite{Campbell2020} outlined a third potential avenue to scaling by using the interaction between polar molecular ions and the phonon modes of a multi-ion Coulomb crystal for quantum logic.
Broadly speaking, this dipole-phonon quantum logic (DPQL) provides a means to initialize, process, and read out quantum information encoded in molecular ion qubits without the need to optically illuminate the molecules (and in some cases without the need for ground-state cooling). 
Together with the use of the direct, electromagnetic dipole-dipole interaction between trapped molecular ions~\cite{Hudson2019}, this could provide a route to a robust and scalable platform for quantum computation, sensing, and communication, including the possibility of microwave-optical transduction \cite{Campbell2020}.

Though the simplicity of DPQL makes it an attractive approach,  it is not without requirements for practical implementation. 
Chief among these is the prerequisite that the qubit molecule possess two opposite-parity states that are energetically spaced by an interval similar to the energy of the trapped-ion phonon modes, i.e.\ $ \lesssim 20$~MHz. 
While the $\ell$- and $K$-doublets of many linear polyatomic and symmetric top molecules satisfy this requirement\cite{Ivan2017}, this criterion is more restrictive for diatomic molecules, where typically only certain $\Omega$-doublets have the necessary structure. 
As the low density of populated rovibrational states found in diatomic molecules makes them attractive candidates for DPQL, identifying molecules with the requisite properties for and understanding their application to DPQL is necessary before experimental implementation.

Here, we study a class of diatomic polar molecules, the cationic alkaline-earth monoxides and monosulfides (MO$^+$ and MS$^+$) and their application to DPQL. 
Through \emph{ab initio} calculations, we find that these molecules possess a large electric dipole moment ($\geq 7$~D) and exhibit $\Lambda$-doublets with $\sim 1$~kHz to $\sim 10$~MHz spacing, making them particularly amenable to DPQL.
Further, since the trapped \textit{atomic} ions used in quantum information experiments are predominantly alkaline-earth atoms, these molecules can be formed \textit{in situ} by simply `leaking' e.g. oxygen into the vacuum chamber~\cite{Goeders2013}.
As DPQL requires no additional lasers, this means many current experimental platforms could easily be extended to this work. 
We follow this discussion with a case study of using DPQL to observe a phonon-mediated interaction between two trapped CaO$^+$ ions. 
We conclude with considerations for practical implementation of this scheme, as well as report on recent progress towards an experimental demonstration of this technique.

\section{Survey of the alkaline-earth monoxides and monosulfides}


\begin{table*}[h]
\resizebox{2\columnwidth}{!}{%
\begin{tabular}{ccccccccccccccc}
\hline

&$A_{\mathrm{SO}}$ &$B_e$ &$\omega_e $&${D_e}$ &PDM &$10^{-3} \cdot T_e$&\multicolumn{4}{c}{$\Lambda$-doublet splitting }&\multicolumn{4}{c}{4~K/300~K Population}\\

&(cm$^{-1}$)&(cm$^{-1}$)&(cm$^{-1}$)&(eV)&(Debye)&(cm$^{-1}$)&\multicolumn{4}{c}{(MHz)}&\multicolumn{4}{c}{(\%)}\\

& & &&&&&$J\!=\!3/2$&$5/2$&$7/2$&$9/2$&$J\!=\!3/2$&$5/2$&$7/2$&$9/2$ \\

\cline{8-11} \cline{12-15}

BeO$^+$ &-117& 1.44&1242&3.8&7.5&9.4 & 3.8&16&42&84&89/2.1 &10/3.1&1/3.9&0/4.6 \\

MgO$^+$ &-130& 0.53 &718&2.3 &8.9&7.3&0.19&0.84&2.1&4.3&54/0.8& 32/1.1&11/1.5&3/1.8\\

CaO$^+$&-130 & 0.37&634&3.3 &8.7&0.7&0.45&1.9&5.0&10&42/0.5&32/0.8&17/1.0&6/1.3\\

SrO$^+$&-147 & 0.31&659&4.2 &7.5&0.4&0.16&0.67&1.7&3.5&0/0.1&0/0.1&0/0.1&0/0.1\\

BaO$^+$&-214 & 0.24&506&2.2 &7.9&1.5&0.089&0.39&0.98&2.0&0/0&0/0&0/0&0/0\\

YbO$^+$ & -132& 0.28& 601&2.2 &7.0&1.0&0.14&0.59&1.5&3.0&33/0.4&30/0.6&20/0.8 &10/1.0\\

RaO$^+$  & -228& 0.20&451&3.3 &7.7&0.3& 0.081&0.35&0.89&1.8&25/0.3&26/0.5&21/0.6 &14/0.8\\

BeS$^+$ & -310 &0.71&875&3.4 &7.6&15.7& 0.11&0.45&1.2&2.3&66/1.2&27/1.8&6/2.3&1/2.8\\

MgS$^+$ & -303&0.25&469&2.0 &9.2&12.9&0.0051&0.022&0.056&0.11&30/0.4&29/0.6&21/0.8&12/1.0\\

CaS$^+$ & -299&0.15&390&4.0 &11.3&5.0&0.0028&0.012&0.031&0.062&19/0.3&22/0.4&20/0.5&16/0.6\\

SrS$^+$  & -316 & 0.12& 423&3.1&8.7&0.3&0.0086&0.036 &0.094&0.19& 16/0.1&20/0.2&19/0.3&16/0.4\\

BaS$^+$  & -273 & 0.08& 291&3.3&9.1&2.5&0.00093&0.0040 &0.010&0.021& 11/0.1&14/0.2&15/0.2&15/0.3\\

YbS$^+$  & -254 & 0.10& 345&2.2&7.8&4.6&0.0013&0.0056 &0.014&0.029& 13/0.1&17/0.2&18/0.3&16/0.3\\

RaS$^+$  & -405 & 0.07& 266&4.5&9.4&2.7&0.00043&0.0018 &0.0047&0.0094& 10/0.1&13/0.1&14/0.2&14/0.2\\

\hline
\end{tabular}}
    \caption{  A list of dipole-phonon quantum logic (DPQL) candidates in electronic state $^2\Pi_{3/2}$. This table includes the spin-orbital coupling constant ($A_{\mathrm{SO}}$), rotational constant ($B_e$), vibrational constant ($\omega_e$), dissociation energy (${D_e}$), permanent dipole moment (PDM), energy interval between two lowest electronic states (${T_e}$), $\Lambda$-doublet splitting and population of several low-lying rotational states. The ground electronic state of all the species in this table is $X^2\Pi_{3/2}$, except for SrO$^+$ and BaO$^+$, whose ground state is $X^2\Sigma^+$ and the first excited state is $A^2\Pi_{3/2}$ .
   }
    \label{table1}

\end{table*}

Cationic alkaline-earth monoxides (see Table~\ref{table1}) play important roles in combustion, atmospheric chemistry, and astrochemistry. They have been the subject of several theoretical studies. 
Electronic structure models predict that the bonding is predominantly ionic, with a lowest-energy electronic configuration that can be formally represented as M$^{2+}$O$^-(2p^5)$. 
The M$^{2+}$ ion is closed shell and the O$^-(2p^5)$ anion may approach with the half-filled orbital pointing towards M$^{2+}$ (2$p\sigma$, resulting in a $^2\Sigma^+$ state) or perpendicular to the internuclear axis (2$p\pi$, resulting in a $^2\Pi$ state). 
For the lighter alkaline-earth ions Be$^{2+}$, Mg$^{2+}$, and Ca$^{2+}$, the electrostatic attraction favors a filled 2$p\sigma$ orbital so that the vacancy resides in 2$p\pi$, yielding an X$^2\Pi$ ground state. 
The configuration with the vacancy in the 2$p\sigma$ orbital produces a low-lying A$^2\Sigma^+$ electronically excited state.
This state ordering is reversed for Sr$^{2+}$ and Ba$^{2+}$ because the interatomic electron-electron repulsion terms become more significant, favoring the approach of the half-filled 2$p\sigma$ orbital, resulting in an X$^2\Sigma^+$ ground state.  
Surprisingly, \emph{ab initio} calculations for RaO$^+$ suggest that the state ordering reverts to X$^2\Pi$ due to relativistic effects.

There are several notable systematic trends for the low-energy states arising from the M$^{2+}$O$^-(2p^5)$ configuration. 
First, as the $^2\Sigma^+$ state involves the direct approach of the half-filled 2$p\sigma$ orbital, the equilibrium distance for the $^2\Sigma^+$ state is smaller than that of the $^2\Pi$ state, regardless of the state energy ordering. 
Second, except for RaO$^+$ where relativistic effects are important, the spin-orbit coupling constant ($A_{\mathrm{SO}}$) for the $^2\Pi$ state predominantly reflects the spin-orbit interaction of O$^-(2p^5)$ where $\zeta_{2p}$=$-118.1$~cm$^{-1}$\cite{Neumark1985}.  
This term is negative (E($\Omega$=3/2)<E($\Omega$=1/2)) as the $^2\Pi$ state is derived from the more than half-filled 2$p\pi^3$ configuration. 
Third, because these low-lying states are derived from the same 2$p^5$ configuration, the interactions between them can be adequately treated using the pure precession model of Van Vleck\cite{Van}. 
The most important interaction, from the perspective of the present model, is the coupling of the $^2\Pi$ and $^2\Sigma^+$ states that gives rise to the energy splitting of the $^2\Pi$ state $\Lambda$-doublets.  
As described below, calculation of the $\Lambda$-doublet splitting requires knowledge of the potential energy curves and molecular constants for both the $^2\Pi$ and $^2\Sigma^+$ states.
The $\Lambda$-doublet components of the $^2\Pi$ state are conventionally labeled as $\mathrm{e}$ for the levels with parity $(-1)^{J-1/2}$  and $\mathrm{f}$ for levels with parity $-(-1)^{J-1/2}$.
When the $^2\Sigma^+$ state is above the $^2\Pi$ state the perturbation pushes the $\mathrm{e}$ component below the $\mathrm{f}$ component for a given value of $J$.
In the present work we are focused on the $\Lambda$-doublet splitting of the $^2\Pi$, $v$=0 level.
If the $^2\Sigma^+$ state is somewhat below the $^2\Pi$, $v$=0 level, the vibrationally-excited levels of the $^2\Sigma^+$ state that are above $^2\Pi$, $v$=0 can still result in the e component being pushed below the f component (see eqn (1) and (2) below).

This overall description of the properties of M$^{2+}$O$^-(2p^5)$ species is also valid for Yb$^{2+}$(4$f^{14}$)O$^-(2p^5)$ and the alkaline-earth monosulfide ions M$^{2+}$S$^-$(3$p^5$), which we also include in Table~\ref{table1}. 
For the latter the $^2\Pi$ spin-orbit splitting is stronger, reflecting the larger S$^-$(3$p^5$) spin-orbit coupling constant of $\zeta_{3p}$=$-321$~cm$^{-1}$\cite{Lineberger1970}.

For the present purpose we focus on species that have a $^2\Pi$ ground state and are primarily concerned with their permanent electric dipole moment (PDM) and the magnitude of the $\Lambda$-doublet splittings of the low-energy rotational levels.
These parameters have not been measured previously and the spectroscopic data needed for theoretical predictions are mostly unavailable (limited spectroscopic data have been reported for CaO$^+$\cite{VANGUNDY2018} and BaO$^+$\cite{Bartlett2015}). 
Consequently, we use \textit{ab initio} electronic structure models to calculate the relevant molecular parameters. 
The results reported here were obtained using the Molpro suite of programs\cite{Werner}.
The low-energy states were predicted using a standard sequence of methods. 
First, a state-averaged complete active space – self-consistent field calculation (SA-CASSCF) was carried out to provide the orbital input for a multi-reference singles and doubles configuration interaction (MRSDCI) calculation.  
The spin-orbit interaction was then included by evaluating the matrix elements of the Breit-Pauli Hamiltonian in the CASSCF eigenbasis.  
All-electron basis sets of augmented valence quadruple zeta quality were used for all of the ions with the exception of RaO$^+$ and YbO$^+$. 
Relativistic effective core potentials for 78 (ECP78MDF) and 28 (ECP28MWB) electrons were used for these metals, respectively.  

Potential energy curves were constructed from single point energies obtained from \textit{ab initio} calculations, with a typical grid spacing of 0.04~$\angstrom$. 
Morse potential energy functions were fitted to these data, providing direct values for the equilibrium distances ($R_e$) and term energies ($T_e$).  
The spin-orbit coupling constants were obtained from the energy difference between the $T_e$ values for the $^2\Pi_{3/2}$ and $^2\Pi_{1/2}$ potential energy curves. 
Harmonic vibrational frequencies ($\omega_e$) were obtained from the fitted Morse parameters using a standard algebraic expression\cite{Herzberg}. 
The results from these calculations are collected in Table~\ref{table1}. 
The body-fixed permanent electric dipole moments and other calculated values determined for the equilibrium bond lengths are also provided in Table~\ref{table1}.
Electronic structure calculations have been reported previously for BeO$^+$\cite{Partridge1986,Ghalila2008}, MgO$^+$\cite{Partridge1986,Bauschlicher1994,Maatouk2011}, CaO$^+$\cite{Partridge1986,Khalil2013,VANGUNDY2018}, SrO$^+$\cite{Partridge1986,Dyke1987}, BaO$^+$\cite{Dyke1987,Bartlett2015}, BeS$^+$\cite{Larbi2010}, and MgS$^+$\cite{CHEN2017} (and references therein).
Overall, the present results are in reasonable agreement with calculations performed at a similar level of theory.  Note that the sign of the spin-orbit coupling constant given for BeO$^+$($X^2\Pi$) in ref.~\citen{Ghalila2008} is in error. 

\begin{figure}
    \centering
    \includegraphics[width=\linewidth]{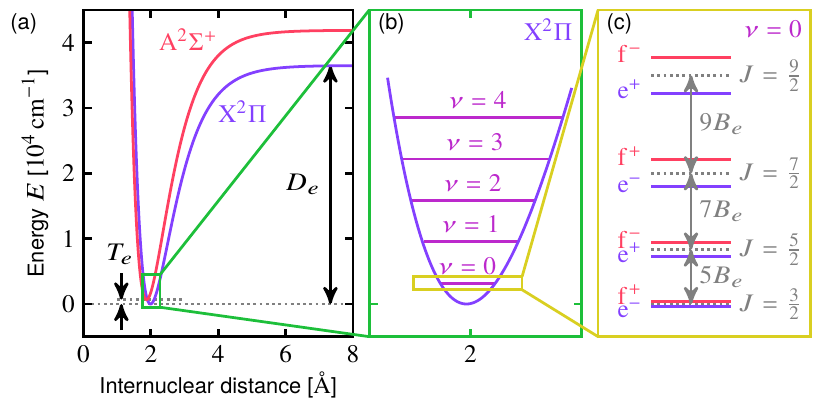}
    \caption{Energy level scheme of CaO$^+$.
    (a) Potential energy surfaces of the two lowest electronic states. (b) The low-lying vibrational states of $X^2\Pi_{3/2}$. (c) The low-lying rotational states in $X^2\Pi_{3/2}$,$v=0$. The $\Lambda$-doublet splittings are not to scale---they are exaggerated for better visibility.}
    \label{level}
\end{figure}

The interactions that result in $\Lambda$-doublet splittings within a $^2\Pi$ state can be incorporated into an effective Hamiltonian~\cite{Mulliken1931} for the rotational motion by means of the constants $p_v$ and $q_v$. 
These parameters can be approximated by the expressions\cite{Zare},

\begin{equation}
    p_v=2\sum_{n',v'}\frac{ \matrixel{n{}^2\Pi v}{A_\mathrm{SO}(R) \vu{L}^+}{n'^2\Sigma^+v'}\matrixel{n'^2\Sigma^+v'}{B(R) \vu{L}^-}{n{}^2\Pi v}}{E_{nv}-E_{n'v'}},
\label{pupsilon}
\end{equation}

\begin{equation}
    q_v=2\sum_{n',v'}\frac{ \Bigr{|}\matrixel{n^2\Pi v}{B(R) \vu{L}^+}{n'^2\Sigma^+v'}\Bigr{|}^2}{E_{nv}-E_{n'v'}},
\label{qupsilon}
\end{equation}
where $A_{\mathrm{SO}}(R)$ is the spin-orbit coupling constant, $B(R)$ is the rotational constant and $\vu{L}^+$/$\vu{L}^-$ are the raising and lowering operators for the electronic orbital angular momentum. 
Evaluation of the matrix elements is simplified by the fact that, near the equilibrium distance both $A_{\mathrm{SO}}(R)$ and $B(R)$ are slowly varying functions of $R$ that can be approximated by their values at $R_e$. 
These `constants' can then be taken outside of the vibrational overlap integrals which then, when squared, are the Franck-Condon factors (FCF).
Finally, with the application of the pure precession model, the matrix elements of the raising and lowering operators are given by $\matrixel{^2\Pi}{\vu{L}^+}{^2\Sigma^+}=\matrixel{^2\Sigma^+}{\vu{L}^-}{^2\Pi}=\sqrt{2} \hbar$.  
The FCFs needed for calculations of the $p_v$ and $q_v$ parameters were obtained for the fitted Morse potentials using the computer code LEVEL 8.0\cite{LeRoy2002}.
Algebraic expressions for the rotational energies of a $^2\Pi$ state, with inclusion of the $\Lambda$-doublet splitting terms, were reported by Mulliken and Christy\cite{Mulliken1931}. 
Eqn~(14) of ref.~\citen{Mulliken1931} was used to calculate the $\Lambda$-doublet splittings listed in Table~\ref{table1}.
Fig.~\ref{level} illustrates the energy level schemes of CaO$^+$ within different energy scales. 

Among these candidate ions, BeO$^+$, BeS$^+$, MgO$^+$, CaO$^+$, YbO$^+$, and RaO$^+$ appear particularly promising. 
These ions have large PDMs, suitable $\Lambda$-doublet splittings, and are easily produced from and co-trapped with commonly laser-cooled atomic ions.
However, the dissociation energies of BeO$^+$, BeS$^+$, MgO$^+$, and YbO$^+$ are less than the energy of the lasers used for Doppler cooling their parent ions: Be$^+$ (3.96~eV), Mg$^+$ (4.42~eV), and Yb$^+$ (3.36~eV).
Therefore, photodissociation from the atomic ion cooling laser could be problematic for these species---although BeO$^+$, for example, could be used with Ca$^+$ (3.12~eV) with added complexity for the trap loading process.
Furthermore, the recent progress on laser cooling of Ra$^+$\cite{Fan2019} makes RaO$^+$ a promising candidate for exploring physics beyond the Standard Model~\cite{Will2017,ACME2018}, since relativistic effects are clearly important in this molecule and it has very small $\Lambda$-doublet splittings. 
Similar to cationic alkaline-earth monoxides, multiply-charged ions such as BO$^{2+}$ also possess $\Lambda$-doublet splittings close to $5$~MHz in the ground rotational state and can also be candidates for DPQL study~\cite{BO2}.

In the remainder of this work, we focus on using CaO$^+$ since it is easily produced from and sympathetically cooled by co-trapped Ca$^+$~\cite{Goeders2013} and its main isotopologue does not possess hyperfine structure.
Unless otherwise noted, we work in the $J\!=\!7/2$ rotational state, which we have calculated to have a blackbody-limited lifetime of $\tau\approx 4$~s at room temperature; this lifetime increases to $\tau \approx 5000$~s at $4$~K. 

\section{Dipole-Phonon Quantum Logic}

At its core, DPQL utilizes the coupling between the motional mode of a Coulomb crystal and the dipole moment of a polar molecular ion to transfer information between a qubit encoded within the internal states of the molecule and a shared phonon mode of the trapped ions. 
Essentially, an oscillating ion experiences an electric field due to the combination of the trapping electric field and repulsion from the other ions.
If the energy interval between the two opposite parity states within the molecule, here assumed to be the two components of a $\Lambda$-doublet, is similar to the energy of the phonon mode, a Jaynes--Cummings type Hamiltonian is realized and the molecule may emit and absorb phonons. 
This interaction provides a complete set of quantum logic operations, including state preparation and measurement of the molecular qubit and atom-molecule and molecule-molecule entanglement. 
In this section we develop the basic building blocks of DQPL and show how they can be used to observe phonon-mediated molecule-molecule interactions.

The dipole-phonon interaction derives from the interaction $H_{dp} = -\mathbf{d}^{(i)} \cdot \mathbf{E}^{(i)}$, where $\mathbf{d}^{(i)} = d \sigma_x^{(i)} \mathbf{\hat{z}}$ is the dipole moment of molecular ion $(i)$, $\mathbf{E}^{(i)}$ is the total electric field at the position of ion $(i)$, $\sigma_x^{(i)} = \ket{\mathrm{e}^{(i)}}\bra{\mathrm{f}^{(i)}}+\ket{\mathrm{f}^{(i)}}\bra{\mathrm{e}^{(i)}}$, and $\mathbf{\hat{z}}$ is the quantization axis, taken to be along the axial direction of a linear Paul trap. 
Assuming only one phonon mode is near the $\Lambda$-doublet splitting $\omega_{\mathrm{mol}}^{(i)}$ of molecule $(i)$ and following the treatment in ref.~\citen{Campbell2020}, the interaction between the molecular qubit state of ion $(i)$ and the axial motion of normal mode $q$ can be described by a spin-boson Hamiltonian:
\begin{equation}
    H_q^{(i)}=\frac{\omega_{\mathrm{mol}}^{(i)}}{2} \sigma_{z}^{(i)}+ \omega_q \left(a_q^{\dagger} a_q + \frac{1}{2}\right)+\frac{g_q^{(i)}}{2} \sigma_{x}^{(i)}\left(a_q^{\dagger}+a_q\right),
\label{eq1}
\end{equation}
where $\hbar = 1$, $\omega_q$ is the secular frequency of the normal mode $q$ along the trap axis, $a_q^\dagger$ and $a_q$ are the creation and annihilation operators for normal mode $q$, and $\sigma_z^{(i)} = \ket{\mathrm{f}^{(i)}}\bra{\mathrm{f}^{(i)}}-\ket{\mathrm{e}^{(i)}}\bra{\mathrm{e}^{(i)}}$. Here, $g_q^{(i)}\equiv d \mathcal{E}_{0,q}^{(i)} = \frac{d}{e}\sqrt{2m^{(i)}\omega_q^3}b_q^{(i)}$ is the vacuum Rabi frequency of the interaction where $\mathcal{E}_{0,q}^{(i)}$ is the electric field amplitude at the position of ion $(i)$ due to a single phonon in normal mode $q$, $m^{(i)}$ is the mass of ion $(i)$, $e$ is the elementary charge, and $b_q^{(i)}$ is the component of the eigenvector of normal mode $q$ at ion $(i)$.
Due to weak coupling, when the normal mode frequency $\omega_q$ is near the resonant frequency of the molecular qubit $\omega_{\mathrm{mol}}$, the rotating wave approximation is justified and eqn~(\ref{eq1}) simplifies to 
\begin{equation}
    H_q^{(i)} \approx \frac{\omega_{\mathrm{mol}}^{(i)}}{2} \sigma_{z}^{(i)}+ \omega_q \left(a_q^{\dagger} a_q + \frac{1}{2}\right)+\frac{g_q^{(i)}}{2}\left(\sigma_{+}^{(i)} a_q+\sigma_{-}^{(i)} a_q^{\dagger}\right),
\label{eq2}
\end{equation}
which, except for the dependence of  $g_q^{(i)}$ on $\omega_q$, is the well-known Jaynes--Cummings model.

\begin{figure}
    \centering
    \includegraphics[width=\linewidth]{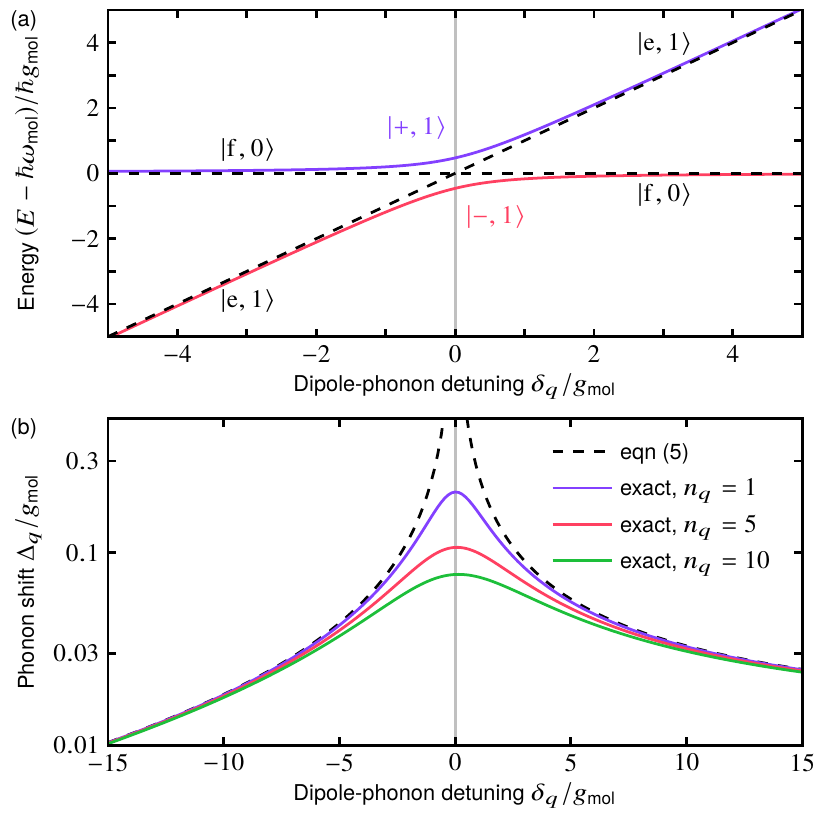}
    \caption{DPQL with one molecule and one atom.
    For the eigenvectors $\ket{+,1}$ and $\ket{-,1}$ of eqn~(\ref{eq2}), the normalized eigenvalues of the energy are plotted in (a) as a function of normalized detuning $\delta_q = \omega_q - \omega_{\mathrm{mol}}$.
    Far away from resonance, these eigenvectors asymptote to the uncoupled eigenvectors $\ket{\mathrm{f},0}$ and $\ket{\mathrm{e},1}$.
    If one starts with normal mode frequency $\omega_q$ far below $\omega_{\mathrm{mol}}$ in the $\ket{\mathrm{f},0}$ state and sweeps $\omega_q$ adiabatically through resonance until it is much larger than $\omega_{\mathrm{mol}}$, the population will adiabatically transfer to the $\ket{\mathrm{e},1}$ state.
    Energies and detunings are normalized by $g_{\mathrm{mol}}$, defined as $\abs{g_q}$ evaluated at $\omega_q = \omega_{\mathrm{mol}}$.
    (b) We calculate the normalized quantum AC Stark shift $\Delta_q$ of normal mode $q$ as a function of normalized detuning from resonance for several values of $n_q$ by numerically solving eqn~(\ref{eq1}) and compare to eqn~(\ref{eq:qACSshift}). 
    In the far-detuned limit, the shift is independent of $n_q$ and eqn~(\ref{eq:qACSshift}) is valid. The shift is asymmetric about $\delta_q = 0$ due to the $\omega_q$-dependence of $g_q$.}
    \label{avoided_crossing}
\end{figure}

\subsection{DPQL with one molecule and one atom}
The dipole-phonon interaction couples the states $\ket{\mathrm{f},n_q}$ and $\ket{\mathrm{e},n_q+1}$, where $\ket{\mathrm{f}}$ and $\ket{\mathrm{e}}$ are the upper and lower $\Lambda$-doublet states, respectively, and $\ket{n_q}$ is the Fock state labeled by the number of phonons $n_q$ in motional mode $q$.
This coupling leads to an avoided crossing when the normal mode frequency $\omega_q$ is near $\omega_{\mathrm{mol}}$, as shown in Fig.~\ref{avoided_crossing} for the case of $\ket{\mathrm{f},0}$ and $\ket{\mathrm{e},1}$.
We define $\delta_q \equiv \omega_q - \omega_{\mathrm{mol}}$ and $g_{\mathrm{mol}}\equiv \abs{g_q} \Bigr\rvert_{\omega_q = \omega_{\mathrm{mol}}} = 2\pi \cdot 58$~kHz for the center-of-mass (COM) mode of an ion chain with a Ca$^+$ ion and a CaO$^+$ molecular ion in the $J\! = \!7/2$ state.
We define the eigenvectors of eqn~(\ref{eq2}) as $\ket{\pm,n_q}$.
As shown in Fig.~\ref{avoided_crossing}, for $\delta_q/g_{\mathrm{mol}} \ll -1$, $\ket{+,1}$ and $\ket{-,1}$ asymptote to $\ket{\mathrm{f},0}$ and $\ket{\mathrm{e},1}$, respectively.
In the limit $\delta_q/g_{\mathrm{mol}} \gg 1$, the situation is reversed with $\ket{+,1}$ converging to $\ket{\mathrm{e},1}$ and $\ket{-,1}$ to $\ket{\mathrm{f},0}$.

Therefore, by adiabatically sweeping the trap potential, $\omega_q$ passes through $\omega_{\mathrm{mol}}$ and induces an adiabatic exchange of energy from the qubit to the motion, with the probability of transition from e.g. $\ket{\mathrm{f},0}$ to $\ket{\mathrm{e},1}$ given as $P\approx1-\exp(-2\pi(g_{\mathrm{mol}}/2)^2/\dot{\omega}_q)$~\cite{Zener1932}.
The desire for high-fidelity transfer, therefore, constrains the rate of the frequency sweep to $\dot{\omega}_q \ll g_{\mathrm{mol}}^2$.
There is an additional constraint on the rate of the frequency sweep, $\dot{\omega}_q \ll \omega_q^2$, in order to retain adiabaticity with respect to normal mode $q$\cite{Schuetzhold2007}.
It has been shown that under this constraint, one can significantly vary the normal mode frequency without significantly changing the Fock state\cite{Poulsen2012}.
This constraint is generally much less stringent than $\dot{\omega}_q \ll g_{\mathrm{mol}}^2$.

This coupling also leads to an effective shift in the normal mode frequency $\omega_q$, which can be understood by examining the energies of states $\ket{+,n_q}$ and $\ket{-,n_q}$.
In the limit $\abs{\delta_q} \gg \abs{g_q}\sqrt{n_q}$, the rotating wave approximation used to write eqn~(\ref{eq2}) is not necessarily valid (if the limit $\abs{\delta_q} \ll \omega_{mol}$ is not maintained) and eqn~(\ref{eq1}) must be used.
Defining the normal mode frequency as the energy required to add a phonon $ E_\pm(n_q+1)-E_\pm(n_q)$, the dipole-phonon interaction leads to shift of the normal frequency by~\cite{Campbell2020}
\begin{equation}
    \Delta_q = E_\pm(n_q+1) - E_\pm(n_q) -\omega_q = \pm \left(\frac{g_q}{2}\right)^2 \frac{2\omega_{\mathrm{mol}}}{\omega_{\mathrm{mol}}^2-\omega_q^2}.
    \label{eq:qACSshift}
\end{equation}
As the $\ket{+,n_q}$ and $\ket{-,n_q}$ states shift the normal mode frequency away from $\omega_q$ in opposite directions, one can determine the molecular state by measuring this frequency shift.
For instance, given a molecule in an unknown internal state in the far-detuned limit, the trap frequency can be brought closer to resonance. For a molecule initially in the $\ket{\mathrm{f}}$ state, the normal mode frequency will increase, while a molecule initially in the $\ket{\mathrm{e}}$ state will lower the motional frequency.
As a quantitative example, for an ion chain with one Ca\textsuperscript{+} ion and one CaO\textsuperscript{+} molecular ion in the $J\!=\!7/2$ state ($\omega_{\mathrm{\textrm{mol}}}=2\pi \cdot 5$~MHz), a center-of-mass (COM) mode frequency $100$~kHz below $\omega_{\mathrm{\textrm{mol}}}/2\pi$ results in a shift of the COM mode frequency of $\pm 7$~kHz, which is easily measured with standard techniques~\cite{Sheridan2011,Goeders2013}.

\par

\subsection{DPQL with two molecules and one atom}

When a second molecule is added to the chain, the molecular dipoles interact directly and via the normal modes of motion.
The direct interaction has been considered in ref.~\citen{Hudson2019} and when $\omega_{\textrm{mol}}$ and $\omega_q$ are comparable it is typically weaker than the phonon-mediated interaction, which we focus on here.
As an illustrative example, we consider a three-ion chain composed of two molecular ions co-trapped with a single atomic ion.
For molecules $(1)$ and $(2)$, the interaction with normal mode $q$ is described by
\begin{equation}
   H_q^{(1,2)} = \frac{1}{2}\left(g_q^{(1)}\sigma_x^{(1)} + g_q^{(2)}\sigma_x^{(2)} \right) \left(a_q^\dagger + a_q\right). \label{eq:2plus1JC}
\end{equation}
This mutual coupling to mode $q$ leads to a coupling of the two dipoles mediated by virtual excitations of the mode.

\begin{figure}[h]
    \centering
    \includegraphics[width=\linewidth]{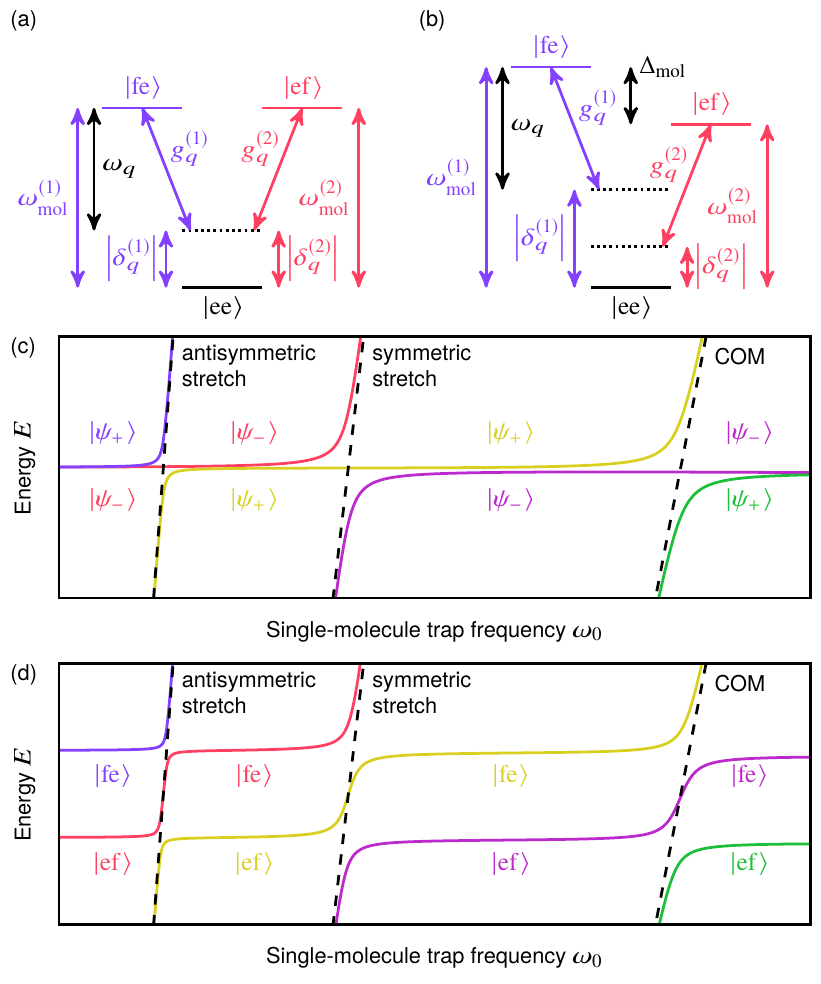}
    \caption{DPQL with two molecules and one atom. (a) Energy levels of several molecular states, illustrating one path in which the mutual coupling of two molecules to the same intermediate state can lead to an indirect coupling of the two molecules via virtual phonons. (b) If the molecular splittings are different, such a two-phonon transition will be off-resonant and the molecules will not be strongly coupled. Also shown are adiabatic potentials for a 3-ion chain with molecular ions at the outer edges for (c) $\omega_{\mathrm{mol}}^{(1)} = \omega_{\mathrm{mol}}^{(2)}$ and (d) $\omega_{\mathrm{mol}}^{(1)} \neq \omega_{\mathrm{mol}}^{(2)}$ as a function of the secular frequency of a single molecular ion. The single-molecular-excitation eigenstates are labeled. These states are of particular importance because they have a one-to-one mapping to phonon states after an adiabatic sweep through resonance. In (c), these states are the Bell states $\ket{\psi^{\pm}} = (\ket{\uparrow \downarrow} \pm \ket{\downarrow \uparrow})/\sqrt{2}$}
    \label{2molecules1atom}
\end{figure}

This virtual-phonon-mediated dipole-dipole (VDD) interaction is similar to a Raman transition, as illustrated in Fig.~\ref{2molecules1atom}, where one path is shown connecting molecular states $\ket{\uparrow\downarrow} \equiv \ket{\mathrm{f}^{(1)},\mathrm{e}^{(2)}}$ and $\ket{\downarrow\uparrow} \equiv \ket{\mathrm{e}^{(1)},\mathrm{f}^{(2)}}$. 
In general for normal mode $q$, there are four paths connecting $\ket{\uparrow\downarrow,n_q}$ and $\ket{\downarrow\uparrow,n_q}$, via intermediate states $\ket{\downarrow\downarrow,n_q+1}$, $\ket{\downarrow\downarrow,n_q-1}$, $\ket{\uparrow\uparrow,n_q-1}$, and $\ket{\uparrow\uparrow,n_q+1}$. The effective Hamiltonian~\cite{James007} for this interaction is then
\begin{multline}
    H_\mathrm{eff} = \sum_{i=1,q}^2 \frac{2\omega_{\mathrm{mol}}^{(i)}}{\left(\omega_{\mathrm{mol}}^{(i)}\right)^2-\omega_q^2}\left(\frac{g_q^{(i)}}{2}\right) ^2\left(a_q^\dagger a_q +\frac{1}{2}\right)\sigma_z^{(i)} \\
    + J_{12}\left(\sigma_+^{(1)}\sigma_-^{(2)}e^{i\left(\omega_{\mathrm{mol}}^{(1)}-\omega_{\mathrm{mol}}^{(2)}\right)t} + \sigma_-^{(1)}\sigma_+^{(2)}e^{-i\left(\omega_{\mathrm{mol}}^{(1)}-\omega_{\mathrm{mol}}^{(2)}\right)t}\right)
\label{eq:VDDhamiltonian}
\end{multline}
where the first term represents the quantum AC Stark shift and the VDD interaction strength is given by
\begin{equation}
    J_{12} = \sum_{q} \frac{g_q^{(1)}    g_q^{(2)}}{4}\left(\frac{\omega_q}{\left(\omega_{\mathrm{mol}}^{(1)}\right)^2-\omega_q^2}+\frac{\omega_q}{\left(\omega_{\mathrm{mol}}^{(2)}\right)^2-\omega_q^2}\right).
    \label{eq:JijDef}
\end{equation}

This VDD coupling facilitates an exchange interaction, similar to the direct dipole-dipole interaction~\cite{Hudson2019}, and has been shown to be significantly stronger than the direct dipole-dipole interaction at long distances\cite{Campbell2020}.
This interaction is also insensitive to phonon number $n_q$.
If the initial state $\ket{\uparrow \downarrow}$ is prepared and allowed to evolve under this interaction, the population will coherently oscillate between $\ket{\uparrow \downarrow}$ and $\ket{\downarrow \uparrow}$.
If $\abs{g_q^{(1)}} \neq \abs{g_q^{(2)}}$ for any mode $q$, it can lead to a differential quantum AC Stark shift $\Delta_{\mathrm{diff}} = \Delta_{\uparrow\downarrow} - \Delta_{\downarrow\uparrow}$ between the $\ket{\uparrow\downarrow}$ and $\ket{\downarrow\uparrow}$ states of the form
\begin{multline}
    \Delta_{\mathrm{diff}} = \sum_q 2\left(n_q+\frac{1}{2}\right) \\
    \times \left[ \left(\frac{g_q^{(1)}}{2}\right)^2 \frac{2\omega_{\mathrm{mol}}^{(1)}}{\left(\omega_{\mathrm{mol}}^{(1)}\right)^2-\omega_q^2}
     - \left(\frac{g_q^{(2)}}{2}\right)^2 \frac{2\omega_{\mathrm{mol}}^{(2)}}{\left(\omega_{\mathrm{mol}}^{(2)}\right)^2-\omega_q^2}\right].
     \label{eq:differentialShift}
\end{multline} 
and the effective Hamiltonian is 
\begin{equation}
    H_\mathrm{eff} = 
    J_{12}\left(\sigma_+^{(1)}\sigma_-^{(2)}e^{i\left(\Delta_{\mathrm{mol}}+\Delta_{\mathrm{diff}}\right)t} + \sigma_-^{(1)}\sigma_+^{(2)}e^{-i\left(\Delta_{\mathrm{mol}}+\Delta_{\mathrm{diff}}\right)t}\right)
    \label{eq:JijSimpleHamiltonian}
\end{equation}
where $\Delta_{\mathrm{mol}} = \omega_{\mathrm{mol}}^{(1)}-\omega_{\mathrm{mol}}^{(2)}$. 
For an initial molecular state $\ket{\psi_{\mathrm{mol}}(0)} = \ket{\uparrow \downarrow}$, the time-dependent probability of flipping to the $\ket{\downarrow \uparrow}$ state $p_{\downarrow \uparrow}(t)$ is
\begin{equation}
    p_{\downarrow \uparrow}(t) = \abs{\braket{\downarrow\uparrow}{\psi(t)}}^2 = \frac{4J_{12}^2}{4J_{12}^2 + \Delta_{\mathrm{tot}}^2} \mathrm{sin}^2\left[\frac{\sqrt{4J_{12}^2 + \Delta_{\mathrm{tot}}^2}}{2}\,\,t\right]
    \label{eq:RabiFloppingVDD}
\end{equation}
where $\Delta_{\mathrm{tot}} = \Delta_{\mathrm{mol}} + \Delta_{\mathrm{diff}}$. 

In what follows, we are specifically interested in two cases.
In the first case, depicted in Fig.~\ref{2molecules1atom}a, $\Delta_{\mathrm{tot}} = 0$ and the population transfers between $\ket{\uparrow \downarrow}$ and $\ket{\downarrow \uparrow}$ with unit amplitude and frequency $2J_{12}$.
The second case, illustrated in Fig.~\ref{2molecules1atom}b, is when $|\Delta_{\mathrm{tot}}| \gg |J_{12}|$.
In this limit, the amplitude of the population transfer is effectively $0$, meaning that the effect of the VDD interaction can largely be ignored.

\subsubsection{State preparation\label{state preparation}}

Often the first step in any operation is to prepare a known state.
To this end, in this section we explore the preparation of $\ket{\downarrow\downarrow}$, a Bell state, and $\ket{\uparrow\downarrow}$ (or $\ket{\downarrow\uparrow}$) in the case that a single Ca\textsuperscript{+} ion is in the center of the chain and surrounded by CaO\textsuperscript{+} molecular ions. A numerical simulation of these techniques is presented in the ESI$\dag$.

\textit{Preparation of $\ket{\downarrow\downarrow}$} --
To prepare the molecular ions in the $\ket{\downarrow\downarrow}$ state, regardless of whether the molecular splittings are identical or not, one can start with all normal mode frequencies below $\omega_{\mathrm{mol}}$.
After ground-state cooling, the combined molecular state $\ket{\psi_{\mathrm{mol}}}$ will be unknown.
By adiabatically increasing the axial confining potential such that the frequency of the two highest-energy modes---in this case the antisymmetric and symmetric stretch modes---sweep through $\omega_{\mathrm{mol}}$, any molecular excitation will be converted to motional excitation of these modes.
By ground-state cooling once again, one prepares the ground molecular state in the ground motional state, $\ket{\downarrow \downarrow,0_q}$.
To verify the state preparation, one can sweep the mode frequencies below $\omega_{\mathrm{mol}}$, apply a global molecular $\pi$ pulse to generate the $\ket{\mathrm{f}\mathrm{f},0_q}$ state, and sweep the frequencies back.
If the state preparation in $\ket{\downarrow \downarrow}$ was successful, this sequence will yield two phonons in the antisymmetric stretch mode.
If detected, these phonons herald preparation in the $\ket{\downarrow \downarrow}$ state.
Alternatively, one can measure the quantum AC Stark shift of one of the normal modes to verify the molecular state.

\textit{Preparation of a Bell state} -- 
Due to the spatial symmetry of the molecules in the ion chain, $\abs{g_q^{(1)}} = \abs{g_q^{(2)}}$ for all modes $q$ and $\Delta_{\mathrm{tot}} = \Delta_{\mathrm{diff}} = 0$. 
Eqn~(\ref{eq:JijSimpleHamiltonian}) then simplifies to 
$H_\mathrm{VDD} = J_{1,2}\sigma_+^{(1)}\sigma_-^{(2)} + \mathrm{H.c.}$
and the eigenstates are then the Bell states $\ket{\psi^{\pm}} = (\ket{\uparrow \downarrow} \pm \ket{\downarrow \uparrow})/\sqrt{2}$.
As shown in Fig.~\ref{2molecules1atom}c, each normal mode couples to only one of these eigenstates.
It is then straightforward to produce the Bell states $\ket{\psi^{\pm}}$ for this case.
Starting with $\ket{\downarrow\downarrow}$ in the ground motional state, one can add a single phonon to the antisymmetric or symmetric stretch mode followed by an adiabatic sweep of both of these mode frequencies across $\omega_{\mathrm{mol}}$ to prepare the $\ket{\psi^{+}}$ or $\ket{\psi^{-}}$ state, respectively.

\textit{Preparation of $\ket{\uparrow \downarrow}$ or $\ket{\downarrow \uparrow}$ } -- It is also possible to use the scheme detailed above to produce $\ket{\uparrow \downarrow}$ or $\ket{\downarrow \uparrow}$ by adding a single phonon in a superposition of the antisymmetric  and symmetric stretch modes  before the sweep of the normal mode frequencies.
In practice, this may be difficult, since the two components of the motional superposition will oscillate at different rates, causing a relative phase difference. 
This relative phase must be at a specific value to produce the desired final state, requiring precise timing of the ramping sequence.

An alternative method to produce the $\ket{\uparrow \downarrow}$ or $\ket{\downarrow \uparrow}$ state is to generate a difference in the molecular splittings of the two molecules, depicted in Fig.~\ref{2molecules1atom}b.
Specifically, if $\abs{\Delta_{\mathrm{tot}}} \gg |J_{12}|$, the effect of the VDD interaction can be ignored.
The eigenstates of eqn~(\ref{eq:JijSimpleHamiltonian}) for this case are simply the basis states $\ket{\uparrow \downarrow}$ and $\ket{\downarrow \uparrow}$.
By starting with $\ket{\downarrow\downarrow}$ in the ground motional state, one can add a single phonon to the antisymmetric or symmetric stretch mode followed by an adiabatic sweep of both of these mode frequencies across $\omega_{\mathrm{mol}}$ to prepare either $\ket{\uparrow \downarrow}$ or $\ket{\downarrow \uparrow}$ in the motional ground state.

One way to generate such a difference ($\omega_{\mathrm{mol}}^{(1)} \neq \omega_{\mathrm{mol}}^{(2)}$) in the molecular splittings of two different molecules (CaO$^+$, in the $X\,{}^2\Pi_{3/2}$ state, with $J\!=\!7/2$) is by taking advantage of the small differential $g$-factor between $\ket{\mathrm{e},J=7/2,m_J=7/2}$ and $\ket{\mathrm{f},J=7/2,m_J=7/2}$, which arises through the differing nuclear rotational-electronic orbit interactions of the $\Lambda$-doublet states with the $A^2\Sigma^+$ states.
Since there is a lack of spectroscopic data for CaO$^+$, we estimate the differential $g$-factor between $\ket{\mathrm{e},J=7/2,m_J}$ and $\ket{\mathrm{f},J=7/2,m_J}$ of CaO$^+$ $X^2\Pi$ as $\delta g \approx 0.8$~kHz/Gauss by applying eqn~(8) in ref.~\citen{Radford1961}.
For two molecular ions separated by $3$~$\mu$m, a magnetic field gradient of $1.3$~T/cm is required to produce a difference of $20$~kHz in the molecular splittings. 
This value of magnetic gradient can be realized by placing a straight wire $30$~$\mu$m away from the molecular ions and driving $1$~A through the wire, similar to conditions discussed in ref.~\citen{Ospelkaus2008}.

\subsubsection{Observing a phonon-mediated dipole-dipole interaction}
The VDD interaction can be observed by preparing $\ket{\psi_{\mathrm{mol}}(0)} = \ket{\uparrow \downarrow}$, allowing the molecules to interact via the exchange interaction, and then reading out the final molecular state $\ket{\psi_{\mathrm{mol}}(t)}$.
As detailed previously, the $\ket{\uparrow \downarrow}$ state can be prepared by generating a difference in molecular splittings $\Delta_{\mathrm{mol}}$ and ramping the trap frequency.
By suppressing the field used to break the degeneracy for state preparation, $\Delta_{\mathrm{mol}}$ can then be set to $0$, effectively turning on the VDD interaction.
After a wait time of $t_\pi = \frac{\pi}{2J_{12}}$, $\ket{\psi_{\mathrm{mol}}(t)} = -i\ket{\downarrow \uparrow}$ and the degeneracy can again be broken by the control field to turn off the VDD interaction and the state measured as discussed in the next section.
As the $\ket{\downarrow\downarrow}$ and $\ket{\uparrow\uparrow}$ states are unaffected, this interaction facilitates an iSWAP gate.
A numerical simulation of this method is presented in Fig.~S3a (ESI$\dag$).

Performing the same experiment with $\Delta_{\mathrm{diff}} \neq 0$, as is generally true if the two molecules do not have spatial symmetry about the center of the ion chain, will cause the population to partially (but not fully) transfer to $\ket{\downarrow \uparrow}$.
Using the control field, complete transfer can be recovered by setting $\Delta_{\mathrm{mol}} = -\Delta_{\mathrm{diff}}$ instead of $\Delta_{\mathrm{mol}} = 0$, cancelling out the differential quantum AC Stark shift and allowing a complete transfer from $\ket{\uparrow\downarrow}$ to $\ket{\downarrow\uparrow}$. 

An alternative method to transfer the population from $\ket{\uparrow \downarrow}$ to $\ket{\downarrow \uparrow}$ using the VDD interaction is to perform an adiabatic sweep of $\Delta_{\mathrm{mol}}$.
One can start again with $\abs{\Delta_{\mathrm{mol}}} \gg |J_{12}|$ to prepare the $\ket{\uparrow \downarrow}$ state.
By slowly ramping the differential splitting from $\Delta_{\mathrm{mol}}$ to $-\Delta_{\mathrm{mol}}$ at speed $\dot{\Delta}$, the population will adiabatically transfer from $\ket{\uparrow \downarrow}$ to $\ket{\downarrow \uparrow}$ with probability of $P=1-\exp(-2\pi J_{12}^2/\dot{\Delta})$.
This method is simulated in Fig.~S3c (ESI$\dag$).
While this method is slower than the first method described, it has the advantage of being insensitive to the differential quantum AC Stark shift $\Delta_{\mathrm{diff}}$ as long as $\abs{\Delta_{\mathrm{diff}}}<\abs{\Delta_{\mathrm{mol}}}$, and does not require precise timing.
Thus, it is not necesary to know the values of $\Delta_{\mathrm{diff}}$ and $J_{12}$ to achieve complete state transfer.
As such, this method may be useful in general cases where the two molecules do not have spatial symmetry in the ion chain or are otherwise nonidentical.

\subsubsection{State detection}
After applying the VDD interaction, a method is needed for reading out the final molecular state $\ket{\psi_{\mathrm{mol}}(t_{\mathrm{final}})} = \alpha \ket{\uparrow \downarrow} + \beta \ket{\downarrow \uparrow}$.
Before beginning the readout procedure, ground-state cooling can remove any potential motional heating that occurred during the sequence.
Following ground-state cooling, the single-molecule trap frequency can be adiabatically increased such that the antisymmetric and symmetric stretch modes sweep through $\omega_{\mathrm{mol}}^{(1)}$ and $\omega_{\mathrm{mol}}^{(2)}$.
For $|\Delta_{\mathrm{mol}}| \gg |J_{12}|$ there is a one-to-one mapping of molecular excitations onto motional excitations and there will be a single phonon in the state
\begin{equation}
    \ket{\psi_{\mathrm{phonon}}} = \beta\ket{1,0} + e^{i\phi}\alpha\ket{0,1}
\end{equation}
where $\ket{n_\mathrm{as},n_\mathrm{ss}}$ denotes the number of phonons in the antisymmetric and symmetric stretch mode and $\phi$ is a relative phase difference that depends on the timing of the ramping sequence.
Red motional sideband transitions of the atomic ion (repeated over many experiments) can be used to probe the phonon occupations ($|\alpha |^2$ and $|\beta |^2$) in each mode.
To keep the phase information, one can utilize simultaneous, phase-coherent sideband operations on the atomic ion in conjunction with precise timing of the sequence.
For the specific case described here with a single atomic ion at the center of the chain, one cannot efficiently use atomic sideband operations on the symmetric stretch mode, since the atomic ion does not participate in this mode.
This shortcoming can be overcome by e.g. chain reordering operations~\cite{Splatt09} or adding a second atomic ion to the chain.

\subsubsection{Single qubit rotations}
Since the VDD interaction is a tunable exchange interaction, it can be utilized to perform an iSWAP gate.
Thus, with the ability to perform qubit-specific, single-qubit rotations DPQL is sufficient for a universal gate set\cite{Schuch03}.
To perform such single-qubit rotations, one can generate a difference in the molecular splittings and apply a direct microwave pulse to the ion chain; the trap electrodes themselves provide a convenient structure to deliver the microwave radiation.
Although this pulse will not be spatially selective, it can select only the frequency of the target molecule.
Control of the Rabi frequency of such a pulse and, if needed, composite pulse sequences can be used to prevent off-resonant excitation of non-target qubits and heating of the motional modes.

\subsection{DPQL with $N$ molecules and $N$ atoms}
Extending this technique to a general number of molecules $N$ may be of particular interest for applications such as quantum simulation and the transduction of quantum information from microwave to optical frequencies\cite{Campbell2020}. 
To allow for efficient state preparation and measurement of these molecular ions, it may be useful to also have $N$ atomic ions in the ion chain. 
With $N$ atoms and $N$ molecules, one can perform state mirroring\cite{KarbachSpin2005} to create a one-to-one mapping of atomic qubit states to molecular qubit states or vice versa.
By utilizing atomic sideband operations, one can map the state of each atomic qubit to the excitation of corresponding normal modes.
A precisely timed adiabatic sweep of these normal mode frequencies through all the molecular qubit resonance frequencies then maps these phonon states onto the molecular qubit states. 


\section{Practical considerations}
In this section we describe progress towards an experimental demonstration of DPQL and examine the effects of motional heating and electric field noise on DPQL fidelity. Furthermore, we present a method to prepare the molecular qubit in a single $m_J$  sublevel.

\subsection{Experiment}
\begin{figure}
    \centering
    \includegraphics[width=\linewidth]{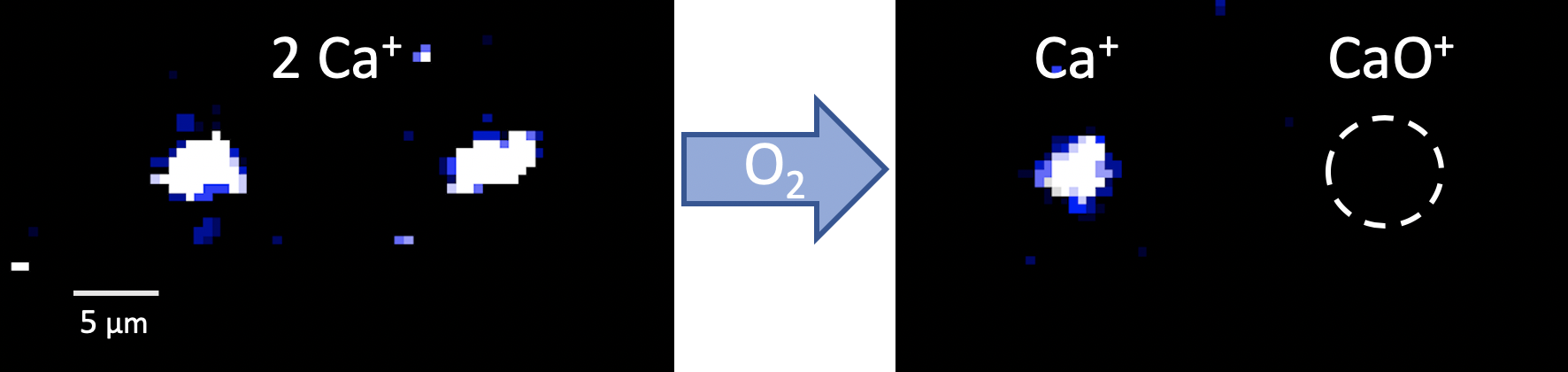}
    \caption{ CaO$^+$ loading protocol. Two Ca$^+$ ions are produced in the trap via resistive heating of a Ca sample and subsequent photo-ionization. O$_2$ gas is introduced into the chamber via a manual leak valve converting one of the Ca$^+$ to CaO$^+$ presumably via the reaction Ca$^+$($^2$P$_{1/2}$) + O$_2 \rightarrow$ CaO$^+$ + O.}
    \label{fig:CaO+}
\end{figure}

\par
To demonstrate these ideas, we have re-purposed an apparatus available to us that was previously used in other trapped molecular ion experiments  and is described in ref.~\citen{Goeders2013}, ~\citen{GoedersThesis}, and ~\citen{RugangoThesis}.
This system can achieve a stable axial secular frequency (COM mode) in the range $2\pi \cdot 180$~kHz~$\leq\omega_{q}\leq 2\pi \cdot 700$~kHz for Ca$^+$. 
Therefore, we leverage the flexibility of molecular systems and choose the $\Lambda$-doublet states of the $J\! =\! 3/2$ manifold, which are separated by $\omega_{\mathrm{mol}} = 2\pi \cdot 450$ kHz, as our qubit. 

To begin, two Ca$^{+}$ are trapped and laser-cooled, see Fig.~\ref{fig:CaO+}.
Oxygen is introduced into the vacuum chamber via a leak valve at a pressure of $5\times 10^{-10}$~Torr until one of the Ca$^+$ reacts to form CaO$^+$. 
Since this reaction is endoergic by $\approx 1.7$~eV, it presumably proceeds as Ca$^+$($^2$P$_{1/2}$) + O$_2$ $\rightarrow$ CaO$^+$ + O.

The experiment is now working to demonstrate state preparation and measurement capabilities for the molecular qubit.
For this, the ions are cooled to the motional ground state via Doppler cooling of Ca$^{+}$ along the $^2$P$_{1/2}\leftarrow$ $^2$S$_{1/2}$ electric dipole transition and sideband cooling along $^2$D$_{5/2}\leftarrow$ $^2$S$_{1/2}$ electric quadrupole transition~\cite{Goeders2013}.
Next, the the DC voltage applied to the trap end caps to form the axial trapping potential will be smoothly increased such that the stretch mode frequency of the ions is scanned from $300$~kHz to $600$~kHz and passes through $\omega_{\textrm{mol}}$.
If the molecule was initially in $\ket{\mathrm{f},0}$ it will transfer to $\ket{\mathrm{e},1}$, and the state can be detected by the presence of a phonon in the symmetric stretch mode.
This phonon is detected by driving a red sideband on the Ca$^+$  $^2$D$_{5/2}$ $\leftarrow$ $^2$S$_{1/2}$ transition and then detecting if the Ca$^+$ ion is in the $^2$D$_{5/2}$ or $^2$S$_{1/2}$.
If the molecule was initially in $\ket{\mathrm{e},0}$ no phonon will be detected and a microwave pulse can be applied to produce $\ket{\mathrm{f},0}$ and the process repeated. 

The fidelity of the transition from $\ket{\mathrm{f},0}$ to $\ket{\mathrm{e},1}$ depends heavily on the sweep rate of the stretch mode frequency, which has upper and lower bounds governed by the requirement for adiabaticity, $\dot{\omega}_q \ll g_q^2$, and the heating rate of the motional mode, respectively. 
To determine the optimum sweep rate, we simulate the evolution of the system with the master equation:
\begin{align}
    \dot{\rho} =& -i[H_s^{(i)},\rho] + \frac{\gamma_s}{2} \left(a^\dagger_s a_s \rho - 2a_s\rho a_s^\dagger + \rho a_s^\dagger a_s\right) \nonumber\\
    &- \frac{\gamma_s}{2} \left(a_s a_s^\dagger \rho - 2a_s^\dagger\rho a_s + \rho a_s a_s^\dagger\right),
\end{align}
where $\rho$ is the molecule-phonon density matrix, $H_s^{(i)}$ is the evolution described in eqn~(\ref{eq2}), and $\gamma_s$ is the stretch mode motional heating rate.
The heating rates of the current trap has been previously measured for the COM mode to be 100 phonons/s for a single ion and 300 phonons/s for two ions~\cite{RugangoThesis}.
The stretch mode heating rate has not been measured, but is expected to be significantly less than this~\cite{King1998}.
This simulation omits contributions of electric field noise heating of the internal states of the molecule.
As the molecular dipole is of order $e a_o$, this heating rate is much smaller than the motional heating rate (see the ESI$\dag$), which is characterized by a transition dipole moment of order 10 $e$nm.

\begin{figure}
    \centering
    \includegraphics[width=\linewidth]{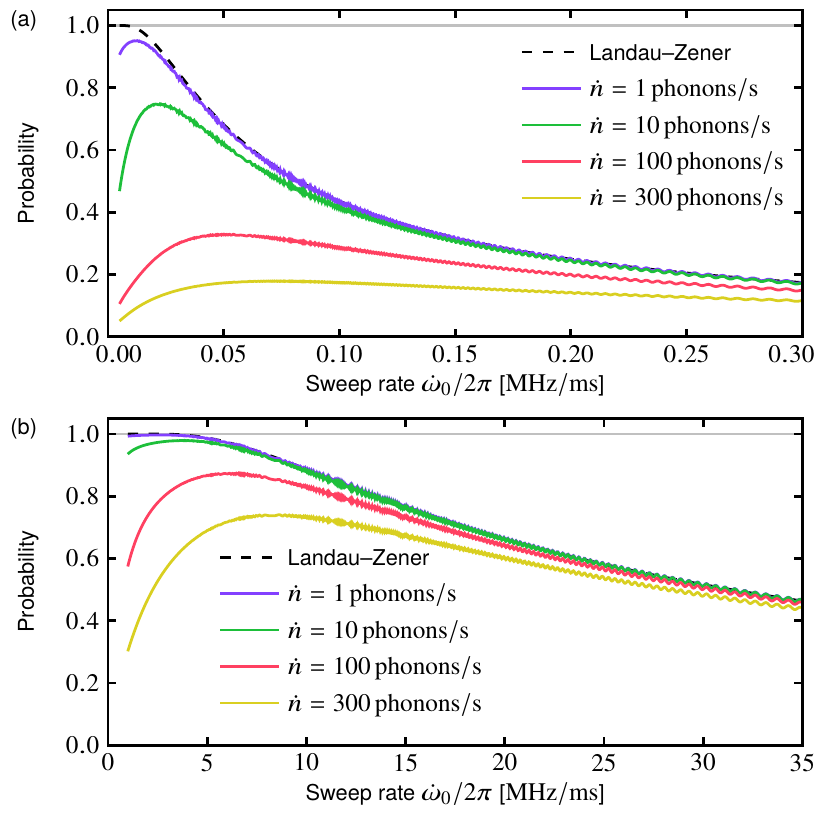}
    \caption{DPQL state-flip probability simulations.
    These plots show the probability of successful state flip from $\ket{\mathrm{f},0}$ to $\ket{\mathrm{e},1}$ after a sweep of the stretch mode frequency for (a) the X$^2\Pi_{3/2}$ $J=3/2$ and (b) the $J=7/2$ states of CaO$^+$ as a function of the sweep rate of the stretch mode frequency for various rates of motional heating. 
    For the $J=3/2$ state the simulated stretch mode frequency was swept from $0.3$ MHz to $0.6$ MHz, while for the $J=7/2$ state it was swept from $3.0$ MHz to $7.0$ MHz. 
    We compare the simulation results to the probability expected from the Landau--Zener formula.
    }
    \label{fig:HeatingRate}
\end{figure}

The results of this simulation are shown in Fig.~\ref{fig:HeatingRate} for both the $J\!=\!3/2$ state, which will be used in the current experiment, and the $J\!=\!7/2$ state explored above. 
The state-flip probability is larger and can be performed with faster sweep rates for the $J\!=\!7/2$ state due to its larger $g_{\mathrm{mol}}$, which scales as $\omega_{\mathrm{mol}}^{3/2}$.
The sweeping time for the high-fidelity transfer is on the order of a millisecond, which is small compared to the blackbody-limited lifetime of the rotational states. 

\subsection{Qubit state preparation}
While the process described above can detect the adiabatic transfer of a molecular excitation to a phonon, without additional operations, it does not resolve the $m_J$ state of the molecule prepared in $\ket{\downarrow}$.
A magnetic field could be used to spectrally resolve the different $m_J$ states through their differential $g$-factor.
However, this slows the process considerably.
Futher, as indicated in Table~\ref{table1}, the energy of $\Lambda$-doublet splitting (and the rotational splitting) of CaO$^+$, X $^2\Pi_{3/2}$ are much smaller than $k_BT$ at room temperature, where $k_B$ is the Boltzmann constant. 
Therefore, all the parities and $m_J$ sublevels are equally populated and the probability of success---defined as the observation of the adiabatic transfer---is small for a single attempt. 
It is therefore desirable, both for efficiency and for use of a qubit, to devise a method to `pump' the molecule into a given $m_J$ using DPQL techniques.

\begin{figure}[t]
\includegraphics[width=\linewidth]{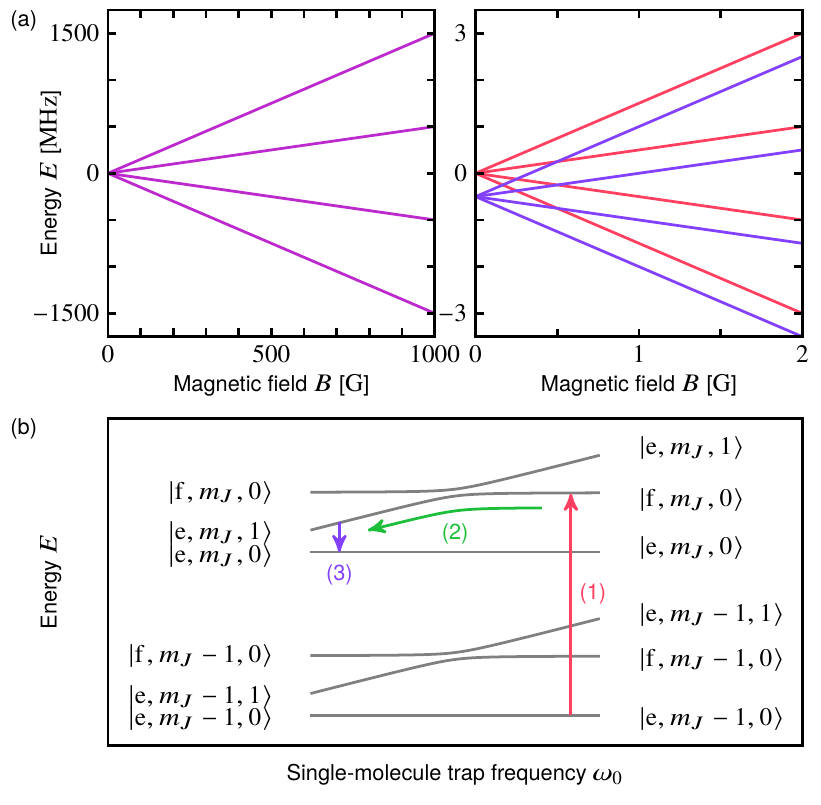}%
\vspace{-5pt}
\caption{\label{shuffling}
Qubit state preparation. (a) Zeeman shift of CaO$^+$, X$^2\Pi_{3/2}$, J=3/2 (left panel). The right panel shows a zoomed part of the left panel with a visible $\Lambda$-doublet splitting. In the right panel, $\mathrm{f}$ and $\mathrm{e}$ parities are indicated by red and blue colors, respectively. (b) The procedure of qubit state preparation. Assume the initial population is cooled to motional ground state of $\mathrm{e}$ parity. (1) Utilize a $\sigma^+$ microwave pulse to move the population from $\ket{\mathrm{e},m_J-1,0}$ to $\ket{\mathrm{f},m_J,0}$. In the ket, there are three notations, which correspond to parity, projection of total angular momentum, and number of phonons, respectively. (2) Adiabatically sweep the normal mode frequency across the $\Lambda$-doublet splitting to transfer all the CaO$^+$ populated in $\ket{\mathrm{f},m_J,0}$ to $\ket{\mathrm{e},m_J,1}$.  (3) Apply sideband cooling to cool CaO$^+$ back to motional ground state. By repeating these steps, CaO$^+$ can be accumulated in stretched state of $\mathrm{e}$ parity ($\ket{\mathrm{e},m_J=3/2,0}$). For this procedure, $800$~G uniform bias magnetic field can be applied to break the degeneracy of sublevels.
}
\end{figure}

In this section, we present a method of preparing CaO$^+$, $^2\Pi_{3/2}$, $J \! = \!3/2$ into its upper stretched state ($m_J \! = \!3/2$) of $\mathrm{e}$ parity\footnote{$J\!=\!3/2$ is used in this example due to its simpler energy structure compared with higher rotational states. 
However, this method can also be applied to other higher rotational states.}. 
Fig.~\ref{shuffling}a shows the Zeeman shift of CaO$^+$, $^2\Pi_{3/2}$, $J$=3/2. 
The right panel of Fig.~\ref{shuffling}a shows the Zeeman shift for smaller magnetic field values, where the $\Lambda$-doublet is noticeable. 
The $\mathrm{e}$ and $\mathrm{f}$ parities are represented by blue and red solid lines, respectively. 
Fig.~\ref{shuffling}b introduces our procedure of state preparation with the assumption that CaO$^+$ has been cooled to its motional ground state of $\mathrm{e}$ parity---$\mathrm{f}$ parity can be addressed by using a microwave pulse. 
(1) A uniform magnetic field is applied along the axial direction to break the degeneracy of the sublevels. 
With a circularly polarized microwave $\pi$-pulse, the populations swap from $\ket{\mathrm{e},m_J-1,0}$ to $\ket{\mathrm{f},m_J,0}$. 
The three notations in the kets represent parity, projection of total angular momentum along quantization axis and number of phonons, respectively.
(2) By adiabatically sweeping the normal mode frequency across the $\Lambda$-doublet splitting, all the populations in $\ket{\mathrm{f},m_J,0}$ can be transferred to $\ket{\mathrm{e},m_J,1}$. 
(3) The quantum state of CaO$^+$ ends in $\ket{\mathrm{e},m_J,0}$ via applying sideband cooling. 
The process is repeated until CaO$^+$ is in the upper stretched state ($m_J \! = \!3/2$) of $\mathrm{e}$ parity, which does not participate in this process.
Due to the fact that the upper stretched state of $\mathrm{e}$ parity is a dark state, the nonequivalent Clebsch-Gordan coefficients of $\sigma^+$ transitions for different $m_J$ does not set any fundamental limit for population transfer.

For the sake of practicality, is is useful to  set the microwave frequency within the range from 500~MHz to a few GHz, so that it can be applied through the trap electrodes.
As shown in Fig.~\ref{shuffling}a,  a magnetic field with amplitude of 800~Gauss will cause a differential Zeeman shift between nearby $m_J$ states up to 800~MHz. 

\section{Summary}
In summary, we have detailed a path towards an experimental demonstration of the basic interactions in dipole-phonon quantum logic (DPQL) for atom-molecule and molecule-molecule systems.
We present and discuss the results of \emph{ab initio} calculations for alkaline-earth monoxide and monosulfide cations.
We find that they appear to be excellent candidates for DPQL as they have nearly ideal rovibronic structure, often do not possess hyperfine structure, and are easily produced in existing trapped ion systems. 
Using a candidate from among these molecules, CaO$^+$, we present early experimental progress and guiding calculations of the expected rates and fidelities for basic DPQL operations with attention to experimental realities, including motional heating and Zeeman substructure. 
We also show, for a chain including two molecules, a scheme to utilize the virtual-phonon-mediated dipole-dipole interaction to perform an iSWAP gate and therefore outline a path to universal quantum computation via DPQL.

\section{Acknowledgements}
This work was supported in part by National Science
Foundation (Grants No. PHY-1255526, No. PHY-1415560, No. PHY-1912555, No. CHE-1900555,
and No. DGE-1650604) and Army Research Office (Grants
No. W911NF-15-1-0121, No. W911NF-14-1-0378,
No. W911NF-13-1-0213 and  W911NF-17-1-0071) grants.


\bibliography{DipolePhonon,chsch}

\bibliographystyle{rsc} 

\end{document}


\maketitle

\section{Effect of electric-field noise on internal states of molecular ions}
\par
There are many sources of electric-field noise in ion traps \cite{PhysRevA.61.063418}: fluctuating electrical fields from injected circuits, varying patch potentials from the trap surface, Johnson noise from the resistance of trap electrodes, and so on. In the DPQL procedure, these noise sources could directly couple to the dipole moments of molecular ions and thus cause transitions between $\Lambda$-doublet splittings and rotational states as well as via ion motion. Here we consider two major forms of electric-field noise\cite{PhysRevA.59.3766}: one is a fluctuating, uniform electrical field $\zeta(t)$; the other is a fluctuating quadrupole field, which perturbs the spring constant of the trap by $\Delta K=\epsilon(t) M \omega_z^2$, where $M$ is ion mass and $\omega_z$ is the axial secular frequency. 
The Hamiltonian under the above noise model can then be written as:
\begin{eqnarray}
H=\frac{P^{2}}{2M}+\frac{1}{2}M(1+\epsilon(t))\omega_{z}^{2}z^2+\zeta(t) q z
\label{eq:hmt}.
\end{eqnarray}
Here $q$ is the charge of ion.
Using first-order time-dependent perturbation theory, we can derive the heating rate near the motional ground state\cite{PhysRevA.56.R1095,PhysRevA.61.045801}:
\begin{eqnarray}
\frac{d}{dt}\langle n \rangle = \frac{\pi\omega_z^2}{4}S_{\epsilon}(2\omega_z) + \frac{q^2}{2M \hbar \omega_z}S_{\zeta}(\omega_z)
\label{eq:zeroheating},
\end{eqnarray}
where $S_{i}(\omega)=\int^{\infty}_{-\infty}d\tau\langle i(t)i(t+\tau)\rangle e^{i\omega \tau}$ (with $i=\epsilon,\zeta$) is the power spectral density  of the fluctuating $\epsilon(t)$ and $\zeta(t)$ respectively and $|n\rangle$ is the motional state. Here, we neglect the higher order term $\langle \epsilon^2(t) \zeta^2(t) \rangle$.
The first term in eqn~(\ref{eq:zeroheating}) is from the quadrupole field fluctuation. 
For a molecular ion that is sympathetically cooled and well compensated from excess micromotion, the coupling between the quadrupole field fluctuation and the molecular ion dipole is small since the electric field noise is close to zero at the RF null. The second term couples the molecular ion dipole through $S_{\zeta}(\omega)$, which results in a spectral energy density $\rho_H(\omega)$
\begin{eqnarray}
\rho_H(\omega)=\epsilon_0 S_{\zeta}(\omega)=\frac{2M \hbar \epsilon_0 \omega_z \Gamma^*}{q^2}
\label{eq:speengdens},
\end{eqnarray}
where $\epsilon_0$ is the permittivity of free space and $\Gamma^*$ is the heating rate caused by $\zeta(t)$. We model the frequency dependence of the electric field noise as a single power law\cite{PhysRevA.91.041402}, $S_{\zeta}(\omega)\propto f^{-0.6}$ over the $\Lambda$-doublet splitting range ($\omega < 2\pi \times 100$ MHz) and broadband noise for higher frequencies. Here we consider the co-trapped Ca$^+$ and CaO$^+$ to have an axial COM frequency of $\omega_z$ = 2$\pi\cdot 450$~kHz and the COM heating rate to be $\Gamma^* = 300$ quanta/s. For $\Lambda$-doublet of $J\!=\!7/2$, electric-field noise results in a transition rate $B_{ef} \rho_H(\omega) \sim 0.001$, where $B_{ef}$ is the Einstein B coefficient for the transition. This transition rate is negligible compared to the thermalization process from blackbody radiation, in which the blackbody-limited life time is 4 s. This result indicates that electric-field noise could affect the DPQL process via motional heating with effective dipole moment of order 10 $e\cdot$nm, but not via directly coupling with dipole moment of molecular ion which is on the order of 0.1 $e \cdot$nm. We also calculate the effect of electric-noise on the rotation state distribution under room temperature using the rate equation model. The results shows that the blackbody-limited lifetime of the states is not affected.

\section{Numerical Simulation}
To demonstrate some two-molecule techniques, we present the results of a numerical simulation of the Hamiltonian
\begin{equation}
    H=\sum_{q,i}\left(
    \frac{\omega_{\mathrm{mol}}^{(i)}}{2} \sigma_{z}^{(i)}+ \omega_q \left(a_q^{\dagger} a_q + \frac{1}{2}\right)+\frac{g_q^{(i)}}{2} \sigma_{x}^{(i)}\left(a_q^{\dagger}+a_q\right)
    \right)
\label{eq1}
\end{equation}
where $\hbar = 1$, $\omega_{\mathrm{mol}}^{(i)}$ is the energy splitting of molecule $(i)$, $\omega_q$ is the secular frequency of the normal mode $q$ along the trap axis, $\sigma_x^{(i)} = \ket{\mathrm{e}^{(i)}}\bra{\mathrm{f}^{(i)}}+\ket{\mathrm{f}^{(i)}}\bra{\mathrm{e}^{(i)}}$, and $\sigma_z^{(i)} = \ket{\mathrm{f}^{(i)}}\bra{\mathrm{f}^{(i)}}-\ket{\mathrm{e}^{(i)}}\bra{\mathrm{e}^{(i)}}$. 
Here, $g_q^{(i)}\equiv d \mathcal{E}_{0,q}^{(i)} = \frac{d}{e}\sqrt{2m^{(i)}\omega_q^3}b_q^{(i)}$ is the vacuum Rabi frequency of the interaction where $\mathcal{E}_{0,q}^{(i)}$ is the electric field amplitude at the position of ion $(i)$ due to a single phonon in normal mode $q$, $m^{(i)}$ is the mass of ion $(i)$, $e$ is the elementary charge, and $b_q^{(i)}$ is the component of the eigenvector of normal mode $q$ at ion $(i)$.
For simplicity, the effect of motional heating is not included in these simulations.

We present the results of numerical simulations of a three-ion chain with two CaO$^+$ molecular ions in the $X\,{}^2\Pi_{3/2}$ state, with $J\!=\!7/2$ on the outside ends of the chain and a single Ca$^+$ atomic ion at the center of the chain. We present simulations of various techniques for state preparation and for demonstrating the virtual-phonon-mediated dipole-dipole interaction.
To demonstrate the results, we plot the "average excitation" of each molecule $(i)$ and each normal mode $q$.
 We define the "average excitation" of molecule $(i)$ to be the probability of measuring the molecule in the $\ket{f^{(i)}}$ state $\abs{\braket{f^{(i)}}{\psi(t)}}^2$.
We define the "average excitation" of normal mode $q$ to be the average phonon number $\expval{n_q} = \bra{\psi(t)}a_q^\dag a_q \ket{\psi(t)}$.

\begin{figure}[htbp]
\centering
\includegraphics[width=\linewidth]{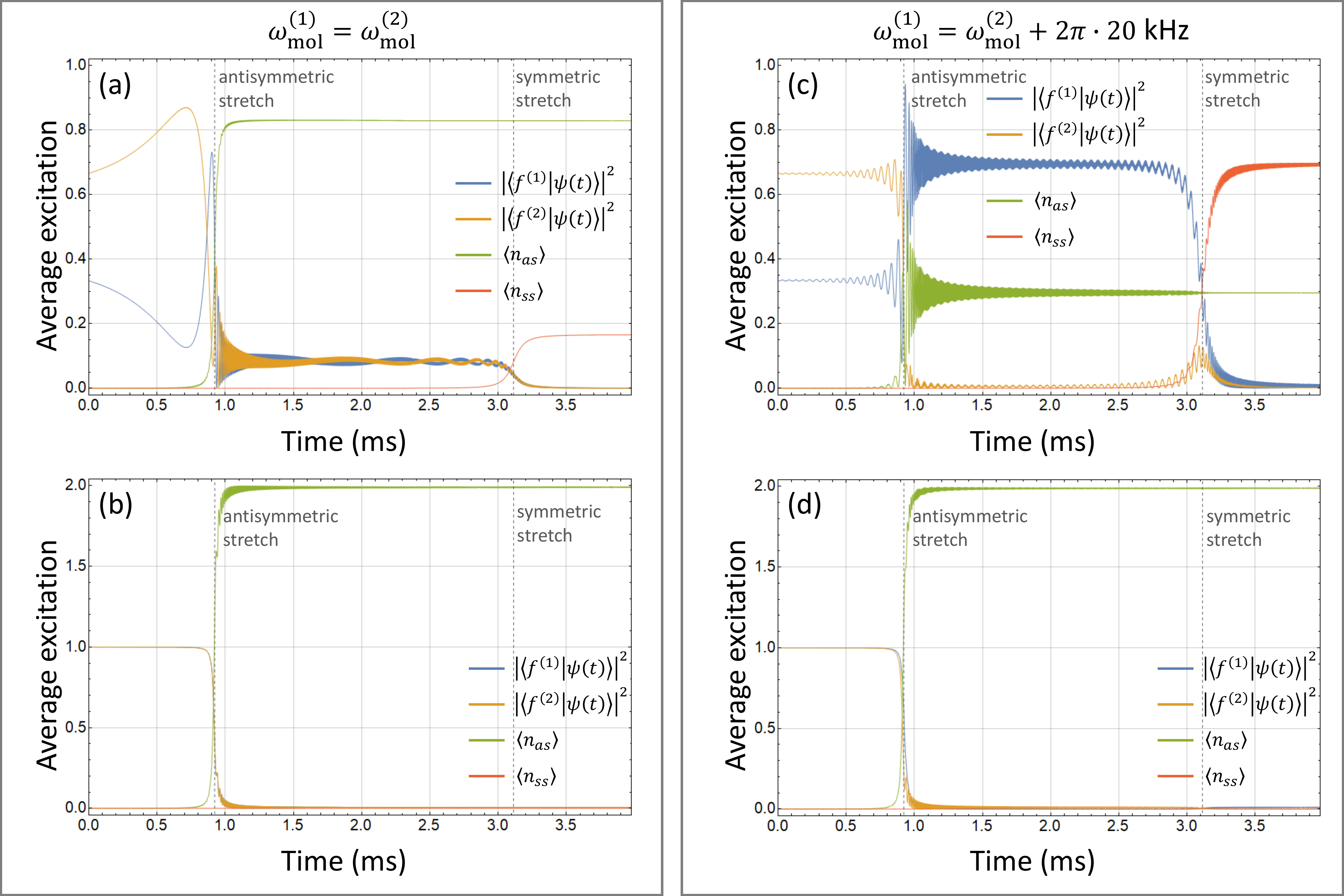}
\caption{Simulation of a frequency sweep to prepare the $\ket{\mathrm{ee}}$ state. We begin with all normal mode frequencies below $\omega_{\text{mol}}^{(1)}$ and $\omega_{\text{mol}}^{(2)}$ in the motional ground state for all normal modes. We then increase the single-molecule trap frequency $\omega_0$ linearly at a rate of $\dot{\omega}_0 = 3.0\times10^9$~rad/s\textsuperscript{2} until both the antisymmetric and symmetric stretch mode frequencies pass through $\omega_{\text{mol}}^{(1)}$ and $\omega_{\text{mol}}^{(2)}$. (a) With equal splittings, we select an arbitrary single-molecular-excitation initial state $\ket{\psi_{\text{mol}}(0)} = \sqrt{\frac{2}{3}}\ket{\mathrm{ef}} + \sqrt{\frac{1}{3}}\,e^{i\pi /4}\ket{\mathrm{fe}}$.
(b) With equal splittings ($\omega_{\text{mol}}^{(1)} = \omega_{\text{mol}}^{(2)}$), we start with $\ket{\psi_{\text{mol}}(0)} = \ket{\mathrm{ff}}$. In this plot, the average excitations for molecules $(1)$ and $(2)$ are overlapping within the thickness of the line.
(c) $\omega_{\text{mol}}^{(1)} = \omega_{\text{mol}}^{(2)} + 2\pi \cdot 20$~kHz and $\ket{\psi_{\text{mol}}(0)} = \sqrt{\frac{2}{3}}\ket{\mathrm{ef}} + \sqrt{\frac{1}{3}}\,e^{i\pi /4}\ket{\mathrm{fe}}$. Far away from resonance, $\abs{\omega_{\text{mol}}^{(1)} - \omega_{\text{mol}}^{(2)}} \gg \abs{J_{12}}$, and the amplitude of oscillation between $\ket{\mathrm{fe}}$ and $\ket{\mathrm{ef}}$ is small. As one of the normal mode frequencies approaches resonance, the coupling strength $J_{12}$ increases, and the previously stated limit is no longer valid. In this regime, there are nontrivial oscillations between $\ket{\mathrm{fe}}$ and $\ket{\mathrm{ef}}$. 
(d) $\omega_{\text{mol}}^{(1)} = \omega_{\text{mol}}^{(2)} + 2\pi \cdot 20$~kHz and $\ket{\psi_{\text{mol}}(0)} = \ket{\mathrm{ff}}$. Again, the average excitations for molecules $(1)$ and $(2)$ are overlapping within the thickness of the line.
In every case shown, sweeping the frequency such that the antisymmetric and symmetric stretch mode frequencies pass through $\omega_{\text{mol}}^{(1)}$ and $\omega_{\text{mol}}^{(2)}$ removes any molecular excitation and prepares the molecular state in $\ket{\mathrm{ee}}$. Additionally, if the initial molecular state is $\ket{\psi_{\text{mol}}(0)} = \ket{\mathrm{ff}}$, such a sweep will yield two phonons in the antisymmetric stretch mode--as this is the first mode to sweep through resonance--regardless of the difference in molecular splittings.
}

\label{fig:S1}
\end{figure}

\begin{figure}[htbp]
\centering
\includegraphics[width=\linewidth]{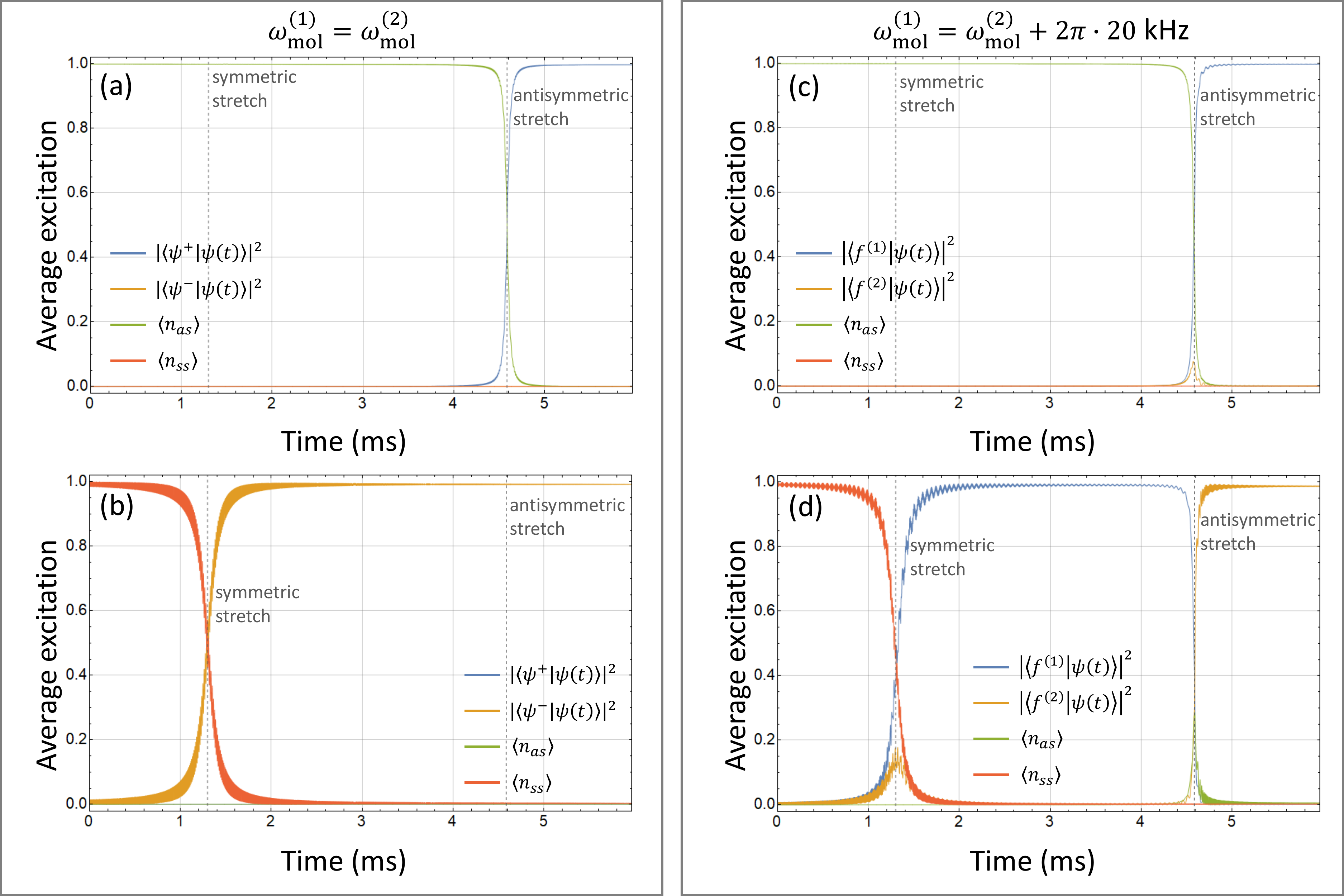}
\caption{
Simulation of molecular state preparation techniques.
For each simulation, the initial molecular state is $\ket{\mathrm{ee}}$ with a single phonon in either the symmetric or antisymmetric stretch mode, and the initial single-molecule trap frequency is tuned such that the symmetric and antisymmetric stretch mode frequencies are above $\omega_{\text{mol}}^{(1)}$ and $\omega_{\text{mol}}^{(2)}$.
The single-molecule trap frequency is decreased at a rate of $\dot{\omega}_0 = 2.0\times10^9$~rad/s\textsuperscript{2} until the antisymmetric and symmetric stretch mode frequencies are below $\omega_{\text{mol}}^{(1)}$ and $\omega_{\text{mol}}^{(2)}$.
(a) With $\omega_{\text{mol}}^{(1)} = \omega_{\text{mol}}^{(2)}$, starting with a single phonon in the antisymmetric stretch mode yields the Bell state $\ket{\psi^+} = \left(\ket{\mathrm{fe}}+\ket{\mathrm{ef}}\right)/\sqrt{2}$.
(b) With $\omega_{\text{mol}}^{(1)} = \omega_{\text{mol}}^{(2)}$, starting with a single phonon in the symmetric stretch mode yields the Bell state $\ket{\psi^-} = \left(\ket{\mathrm{fe}}-\ket{\mathrm{ef}}\right)/\sqrt{2}$.
(c) With $\omega_{\text{mol}}^{(1)} = \omega_{\text{mol}}^{(2)} + 2\pi \cdot 20$~kHz, starting with a single phonon in the antisymmetric stretch mode yields $\ket{\mathrm{fe}}$.
(d) With $\omega_{\text{mol}}^{(1)} = \omega_{\text{mol}}^{(2)} + 2\pi \cdot 20$~kHz, starting with a single phonon in the symmetric stretch mode yields $\ket{\mathrm{ef}}$.
}
\label{fig:S2}
\end{figure}

\begin{figure}[htbp]
\centering
\includegraphics[width=\linewidth]{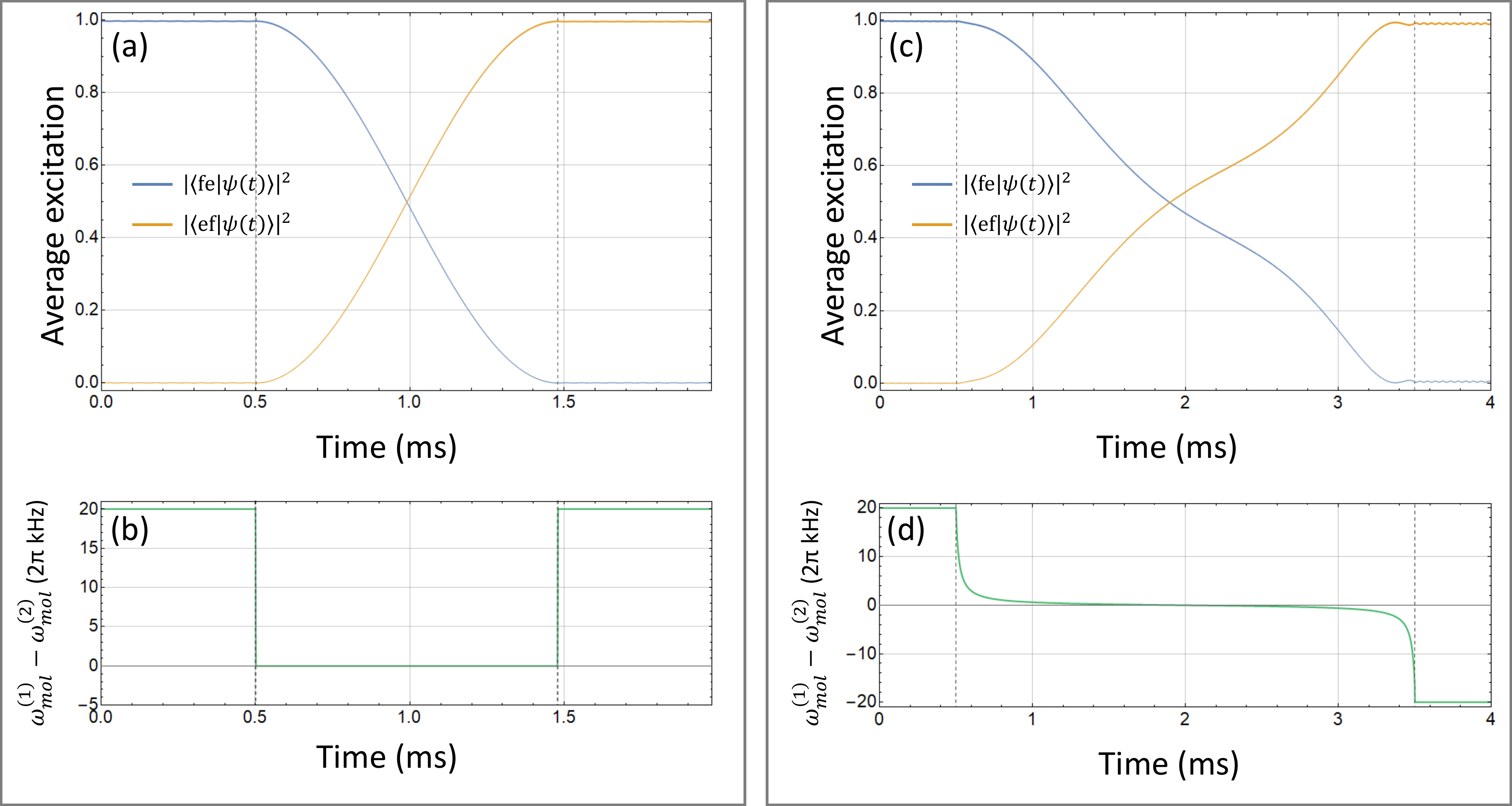}
\caption{Two different methods to transfer the population from $\ket{\mathrm{fe}}$ to $\ket{\mathrm{ef}}$. In the first method, depicted in (a) and (b), we start with the molecular splittings different by $2\pi \cdot 20$~kHz. To begin the transfer, we quickly set this difference to zero, allowing the population to coherently oscillate between $\ket{\mathrm{fe}}$ and $\ket{\mathrm{ef}}$. After waiting $t_\pi = \frac{\pi}{2J_{12}}$, where $J_{12}$ is the virtual-phonon-mediated dipole-dipole interaction strength, we restore the difference in molecular splittings to stop this oscillation. The average molecular excitation of molecules $(1)$ and $(2)$ are shown in (a) and the difference in molecular splittings as a function of time is shown in (b).
For the second method, shown in (c) and (d), we adiabatically vary the difference in molecular splitting. To achieve a high probability transfer with a linear sweep of the molecular splitting difference, it is necessary for the sweep of the molecular splitting difference $\abs{\dot{\Delta}} \ll J_{12}^2$. To dramatically decrease the time needed to perform an adiabatic sweep, we change the molecular splitting difference in a nonlinear fashion, with ramping speed $\dot{\Delta}(t)$. The ramp is faster for larger differences in molecular splittings. For all times, we maintain the limit $\abs{\dot{\Delta}(t)} \ll J_{12}^2 + \left(\frac{1}{2}(\omega_{\text{mol}}^{(1)} - \omega_{\text{mol}}^{(2)})\right)^2$.
}
\label{fig:S3}
\end{figure}

\bibliography{DipolePhonon,chsch}